\documentclass[11pt,a4paper]{article}
\pdfoutput=1
\usepackage{jheppub}
\usepackage{amsmath}
\usepackage{epsfig}
\usepackage{amssymb}
\usepackage{graphics}
\usepackage[active]{srcltx}
\usepackage{epstopdf}
\usepackage{pdfsync}
\usepackage{shuffle}
\usepackage{slashed}
\usepackage{soul}

\usepackage{tikz}
\usetikzlibrary{decorations.pathmorphing}
\usetikzlibrary{arrows.meta}

\def\chalf{{\mathcal{C}}_{\tiny\mbox{half~}}}

\title{On the Seven-loop Renormalization of Gross-Neveu Model}
\author[a]{Rijun Huang}
\author[b]{, Qingjun Jin}
\author[b]{and Yi Li}

\affiliation[a]{Nanjing Key Laboratory of Particle Physics and Astrophysics, School of Physics and Technology, Nanjing Normal University,  No.1 Wenyuan Road, Nanjing 210046, P.R.China}
\affiliation[b]{Graduate School of China Academy of Engineering Physics, No. 10 Xibeiwang East Road, Haidian District, Beijing, 100193, P.R.China }

\emailAdd{huang@njnu.edu.cn}
\emailAdd{qjin@gscaep.ac.cn}
\emailAdd{yili@gscaep.ac.cn}

\abstract{The presence of an infinite number of marginal four-fermion operators is a key characteristic of the two-dimensional Gross-Neveu model. In this study, we investigate the structure of UV divergences in this model, and by symmetry argument we found that the renormalizability only requires a subset of evanescent operators. We perform a 7-loop renormalization computation of beta function for the corresponding evanescent operator, and confirm its non-trivial contribution to UV divergences. By integrating infrared rearrangement, dimensional shifting, and large momentum expansion techniques, we systematically reduce the two-dimensional tensor integrals in the four-fermion correlation functions into four-dimensional scalar integrals. These scalar integrals are subsequently evaluated using the graphical function method, which marks the first application of the method to models with fermionic fields. Our result represents the first time that beta functions have been computed analytically beyond 5-loop in a model with spinning particles.}


\begin{document}
\maketitle \flushbottom

\section{Introduction}
\label{sec:introduction}


Since its initial proposal \cite{Gross:1974jv}, the Gross-Neveu model has spurred extensive research into its fascinating phenomena within quantum field theory. This model is a fermion field theory with four-fermion interactions, and perturbatively renormalizable in two-dimensional spacetime. The Gross-Neveu model is itself of intrinsic importance. It shares many similarities to the quantum chromodynamics (QCD) such as the asymptotic freedom and the large-$N$ expansion, yet it provides a simpler testing ground for exploring fundamental concepts in quantum field theory, for instance the existence of an exact S-matrix \cite{Zamolodchikov:1978xm}, the dynamical chiral symmetry breaking \cite{Witten:1978qu}, {\sl etc}. It is also found that the two-dimensional Gross-Neveu model is intimately connected to other conformal field theory (CFT) in dimensions greater than two, and this fact is supported by the example that in the large-$N$ limit, the Gross-Neveu model and the four-dimensional Gross-Neveu-Yukawa model become equivalent at their renormalization group fixed points \cite{Moshe:2003xn}. Researches also reveal the interdisciplinary connections between this model and the Ads/CFT duality theory \cite{Giombi:2016ejx}. Besides the theoretical interests, the Gross-Neveu model also has applications to condensed matter physics \cite{Shankar:2017zag}, and we suggest the readers refer to more specific literatures.

For the topic we are considering in this paper, the renormalization computation of the Gross-Neveu model has evolved continuously over nearly half a century since its inception, along with many companion theories which also interacting through four-fermion vertex in two-dimensional spacetime. The original $O(N)$ Gross-Neveu model contains the four-fermion vertex $(\bar{\Psi}_i\Psi^i)^2$, where $\Psi_i,i=1,\ldots,N$ is the $N$-component fermion field  \cite{Gross:1974jv}. A changing of the interaction vertex to $(\bar{\Psi}_i\gamma_\mu\Psi^i)^2$ defines the $U(N)$ Thirring model, while it is known that the $N=1$ Gross-Neveu model is equivalent to the $N=1$ Thirring model. A non-trivial generalization of Thirring model is the non-Abelian Thirring model (NATM) in which the interaction vertex is generalized to $(\bar{\Psi}_i\gamma_\mu T^{\alpha}\Psi^i)^2$, where $T^\alpha$ is the group structure constant of $SU(N)$ group \cite{Dashen:1973nhu,Dashen:1974hp}. The NATM is also called the chiral Gross-Neveu model since using Fierz identity the vertex can be related to $(\bar{\Psi}\Psi)^2-(\bar{\Psi}\gamma^5\Psi)^2$ when $N=1$. Another generalization of Thirring model is the $U(N)$ generalized Thirring model (GTM) with both $(\bar{\Psi}_i\Psi^i)^2$ and  $(\bar{\Psi}_i\gamma_\mu T^{\alpha}\Psi^i)^2$ vertices, which is a combination of Gross-Neveu model and NATM \cite{Bondi:1988fp,Bondi:1989nq}. Note that each theory could have possible extensions to other group structures. These theories resemble each other, and advancements of computational techniques are often shared among them, to push them together to a higher order progress. The 2-loop beta function of $O(N)$ Gross-Neveu model is computed in \cite{Wetzel:1984nw}, about ten years after the 1-loop computation in the original paper. The 2-loop renormalization computations of NATM and GTM appear about five years later \cite{Destri:1988vb,Bondi:1988fp,Bondi:1989nq}. In the above mentioned papers, the computations have been done in a Lagrangian formulation with auxiliary field. Note that in \cite{Bondi:1988fp,Bondi:1989nq}, the influence of evanescent operators has been noticed and discussed, revealing its non-trivial impact in the dimensional regularization, which would become more important in high loop renormalization computation. Advances in techniques have boosted the computation to 3-loop order one year later, leading to the 3-loop beta function of Gross-Neveu model \cite{Gracey:1990sx,Tracas:1989hi,Tracas:1990wc,Gracey:1991vy,Luperini:1991sv}. Note that in \cite{Gracey:1990sx,Gracey:1991vy}, the computations have been done in a Lagrangian formulation with four-fermion interaction. Furthermore, a recurrent relation for evaluating the infinite part of certain Feynman integrals has been proposed to compute the highest-$N$ term of the 4 or 5-loop beta function \cite{Tracas:1990kw}. Meanwhile, the 3-loop beta function of NATM is only successfully computed about ten years later \cite{Bennett:1999he}. During these ten years, the evanescent operators have become an important consideration. Techniques have been proposed to deal with the general four-fermion vertex \cite{Vasiliev:1995qj}, and the concept of multiplicative renormalizability due to evanescent operators has been investigated in the computation of 3-loop beta function of Gross-Neveu model \cite{Vasiliev:1996rd,Vasiliev:1996nx,Vasiliev:1997sk}. At 4-loop order and higher, the evanescent operator becomes a more involved and unavoidable problem, this is discussed in the 4-loop field renormalization computation of NATM \cite{Ali:2001he}. Alternatively, the 4-loop beta function of NATM for $SU(N)$, $SO(N)$ and $SP(N)$ groups has been analyzed by applying operator product expansion (OPE) method \cite{Ludwig:2002fu}. For the Gross-Neveu model, however, it has been waited for about twenty years after the 3-loop result, that the 4-loop mass anomalous dimension of $SU(N)$ Gross-Neveu model has been computed \cite{Gracey:2008mf}, and another eight years for the successful computation of the complete 4-loop beta function \cite{Gracey:2016mio}.


The different impact of the evanescent operators on integer dimension and the dimensional regularization is a general property for four-fermion operators \cite{Dugan:1990df}, and its explicit effect depends on the regularization schemes \cite{Herrlich:1994kh}. This phenomenon also appears for gluonic operators in Yang-Mills theory \cite{Jin:2022ivc,Jin:2022qjc}. In general, the evanescent operators in dimensional regularization will give a vanishing result at tree level, but contribute non-trivially to higher loop diagrams. For two-dimensional Gross-Neveu model, unique renormalization structure emerges due to the evanescent operators. It is noticed very early that, in integer dimension only $g_0,g_1,g_2$ couplings have non-zero contributions while in dimensional regularization, the infinite couplings $g_n$'s will contribute \cite{Bondi:1989nq}. Here $g_n$ is the coupling constant of the four-fermion vertex $O_n\sim (\bar{\Psi}\Gamma^{(n)}\Psi)^2$, where $\Gamma^{(n)}$ is a linear combination of the totally anti-symmetrized products of $n$ gamma matrices. The operators $O_n$, $n\geq 2$ are evanescent operators, since they vanish in two-dimensional spacetime but leave imprints in $D=(2-2\epsilon)$ dimensions. Previous researches show that, the $O_3$ \cite{Vasiliev:1996rd,Vasiliev:1996nx} and $O_4$-type \cite{Gracey:2016mio} UV divergences first appear at 3-loop and 4-loop respectively, yet the operators like $O_1,O_2$ remain absent even at 3 and 4-loops, defying the naive power-counting expectation that $O_n$-type divergence will appear at $n$-loop order. A question naturally arises, especially for the high loop renormalization, that to what extend can we ensure the contribution of $O_n$ operator to the $n$-loop renormalization? From a parity symmetry argument, we resolve this question by proposing the {\sl $\chalf$-symmetry} for the Gross-Neveu model, and confirm the observation that only a selective type of evanescent operators will contribute to the same loop order UV divergence. This symmetry-driven selection rule is a mechanism previously unrecognized in the studies of Gross-Neveu model symmetries. In a word, the selection rule says that only the operators $O_n$ satisfying $(n\mod 4)\equiv 0,3$ will contribute to the $n$-loop divergence. This explains the absence of $O_1,O_2$ at 3 and 4-loops, while the $O_3,O_4$ do contribute appropriately. The absence of $O_5,O_6$ can be checked directly by Feynman diagrams, and the next non-trivial operator is $O_7$. Hence through the symmetry analysis, we reveal for the first time the anomalous emergence mechanism of the $O_7$-type operator at the 7-loop perturbative order. To eliminate divergences, a gauge-invariant counterterm $g_7 O_7$ must be introduced into the theoretical framework, leading to a corresponding correction term in the Lagrangian. To confirm the contribution of evanescent operator $O_7$, we need to compute the 7-loop beta function $\beta(g_7)$.

The possibility of computing the 7-loop beta function $\beta(g_7)$ relies on many available techniques. The most crucial is the graphical function method \cite{Schnetz:2013hqa,Golz:2015rea,Borinsky:2021gkd,Schnetz:2024qqt,HP} which allows one to compute very high loop scalar integrals in $D\geq 4$ even-integer dimensions and in dimensional regularization. The ability of computing the 7-loop integrals of quartic scalar theory \cite{Schnetz:2022nsc} is especially useful for our research. The other techniques include the large momentum expansion for extracting UV divergence \cite{Huang:2024hsn}, which allows one to transform the four-point function to a two-point function and extracts the UV divergences that are necessary for determining the renormalization. This greatly reduces the complexity of computation, and make all the integrals of $\beta(g_7)$ calculable. As well as other traditional techniques such as the organization of Feynman diagrams via permutation groups \cite{Vasiliev:1997sk}, the transformation of tensor integral to higher dimensional scalar integral via dimensional shifting relation \cite{Tarasov:1996br,Tarasov:1997kx}, {\sl etc}. By reexamining and applying these techniques in the renormalization computation of Gross-Neveu model, we have successfully reproduced $\beta(g_3),\beta(g_4)$ which are consistent with other literatures, and most importantly the 7-loop result $\beta(g_7)$, which to our knowledge is the first beta function computed analytically beyond 5-loop in a model with spinning particles.

This paper is organized as follows. In \S\ref{sec:theory}, we provide a quick review on the Gross-Neveu model, the evanescent operators and the auxiliary field formulation, and propose the {$\chalf$-symmetry } for Gross-Neveu model. In \S\ref{sec:7loop} we illustrate various techniques for computing 7-loop beta function $\beta(g_7)$, and the result is presented in \S\ref{subsec:7loopBeta}. Conclusion and discussion are presented in \S\ref{sec:conclusion}.

\section{The Gross-Neveu model and its UV divergence structure}
\label{sec:theory}

In this section, we firstly review the Gross-Neveu model and its auxiliary field formulation. Then by presenting a parity symmetry based argument, we propose the {$\chalf$-symmetry } for Gross-Neveu model.

\subsection{The Gross-Neveu model and the four-fermion vertices}
\label{subsec:Lagrangian}

The $SU(N)$ Gross-Neveu model is a field theory with $N$-component Dirac fermion $\Psi^i$, $i=1,2,\ldots,N$ interacting through four-fermion vertex in two-dimensional spacetime. Its Lagrangian is given by,
\begin{equation}
\mathcal{L}_{\tiny\mbox{GN}}=\bar{\Psi}_i(i\slashed{\partial}-M)\Psi^i-\sum_{n=0}^{\infty}g_n O_n~.
\end{equation}
The infinite vertices $O_n$'s are four-fermion interactions preserving $U(N)$ symmetry, defined as,
\begin{equation}
O_n:=\frac{1}{2(n!)}\left(\bar{\Psi}_i\gamma_{\mu_1\cdots\mu_n}\Psi^i\right)
\left(\bar{\Psi}_j\gamma^{\mu_1\cdots\mu_n}\Psi^j\right)~,
\end{equation}
where $\gamma^{\mu_1\cdots\mu_n} := \gamma^{[\mu_1} \gamma^{\mu_2} \cdots \gamma^{\mu_n ]}$ denotes the anti-symmetrization on the indices. In the original form of the Gross-Neveu model \cite{Gross:1974jv}, only the $O_0$-type interaction was considered. However, the operators $O_n$ with $n\ge 1$ are also marginal operators in two-dimension, and they are essential for ensuring renormalizability if one works in dimensional regularization scheme. The operators $O_n$ with $n\geq 2$ are evanescent operators. It was found that $O_3$ \cite{Vasiliev:1996rd,Vasiliev:1996nx} and $O_4$-type \cite{Gracey:2016mio} UV divergences first appear at 3-loop and 4-loop, respectively.

The UV counterterms of four-fermion correlation function has the following structure,
\begin{equation}
\Bigl\langle \Psi^{i_1\alpha_1}(p_1)\bar{\Psi}_{i_2\alpha_2}(p_2)\Psi^{i_3\alpha_3}(p_3)\bar{\Psi}_{i_4\alpha_4}(p_4) \Bigr\rangle
\sim \sum_{n=0}^{\infty}\frac{G_{n}(\epsilon)}{n!}
\left(\left(\gamma^{\mu_1\cdots \mu_n}\right)^{\alpha_1}_{~~\alpha_2}\left(\gamma_{\mu_1\cdots \mu_n}\right)^{\alpha_3}_{~~\alpha_4}\delta_{i_1}^{i_2}\delta_{i_3}^{i_4}
-\bigl(2\leftrightarrow 4\bigr)\right)~,\nonumber
\end{equation}
which gets contributions from all possible $O_n$ vertices, and $G_n(\epsilon)$ is the coefficient of $O_n$-type UV divergences. The $Z$-factor of $g_n$ can be extracted from $G_n(\epsilon)$. Following \cite{Gracey:2016mio}, we focus on the renormalization group flow around,
\begin{equation}
g_0=g~~~,~~~g_i=0~~,~~i=1,2,\ldots
\end{equation}
This means classically, $g_0$ is the only non-vanishing coupling constant, while all other $g_i$'s vanish\footnote{Throughout this paper, we use $\{{\rm{g}}\}/g_0=0$ to represent all $g_i=0$ for $i=1,2,\ldots,\infty$, except $g_0=g\ne 0$.}. The coupling $g_i$ may receive non-zero quantum corrections due to the renormalization if the four-fermion correlation functions produce a $O_i$-type UV divergence. To be more specific, the bare coupling constants $g_i^{{\tiny\mbox{bare}}}$ are non-zero\footnote{Since $Z_i$ usually contains $\frac{1}{g_i}$ poles, in eqn.\eqref{eqn:g-i-bare} $Z_i g_i$ may not vanish even when all $\{{\rm{g}}\}/g_0=0$.},
\begin{equation}\label{eqn:g-i-bare}
g_i^{{\tiny\mbox{bare}}}\Bigr|_{\{{\rm{g}}\}/g_0=0}=
Z_i\widetilde{\mu}^{2\epsilon}g_i\Bigr|_{\{{\rm{g}}\}/g_0=0}\ne 0~,
\end{equation}
so that the corresponding counterterm can cancel the $O_i$-type UV divergence.

When $\{{\rm{g}}\}/g_0=0$, $g_n^{{\tiny\mbox{bare}}}$ starts to appear at $n$-loop \cite{Gracey:2016mio}. To analyze the behaviors of UV divergence, let us first start with four-fermion Feynman diagrams with only $O_0$-vertices, in which case the gamma matrices are only carried by fermion propagators. Considering a $L$-loop diagram, it has $2L$ fermion propagators, so there will be $2L$ gamma matrices in the numerator. Some of the gamma matrices will be contracted, therefore the diagram produces $O_n$-type divergences with $n\le L$. Consequently, $O_n$-type counterterms are required for the renormalization starting at $n$-loop, which means
\begin{equation}\label{eqn:gi-order}
Z_n g_n= \mathcal{O}(g^{n+1})~.
\end{equation}
For four-fermion Feynman diagrams with generic vertices, the $O_n$-vertices also carry gamma matrices. Considering a $(V-1)$-loop diagram with $V$ four-vertices corresponding to $O_{n_1},\cdots ,O_{n_V}$. The propagators still produce $2(V-1)$ gamma matrices, and an $O_{n_i}$ vertex contains $2n_i$ gamma matrices. So the diagram produces at most $O_{n_{\tiny\mbox{max}}}$-type divergences, with the number $n_{\tiny\mbox{max}}$ constrained by,
\begin{equation}
n_{\tiny\mbox{max}}=V-1+\sum_{i=1}^V n_i=-1+\sum_{i=1}^V (n_i+1)~.
\end{equation}
Using eqn.\eqref{eqn:gi-order}, the coupling constants of the diagram is of the order,
\begin{equation}
\prod_{i=1}^VZ_{n_i}g_{n_i}=\mathcal{O}(g^{\sum_{i=1}^V (n_i+1)})=\mathcal{O}(g^{n_{\tiny\mbox{max}}+1})~.
\end{equation}
Hence the $O_{n_{\tiny\mbox{max}}}$-type counterterms of the order $\mathcal{O}(g^{n_{\tiny\mbox{max}}+1})$ are needed to cancel this divergence. This means eqn.\eqref{eqn:gi-order} holds for all four-fermion diagrams with generic four-fermion vertices.

\subsection{The auxiliary field formulation}
\label{subsec:auxiliaryField}

For the cases $n=3,4$, it was verified by direct computations \cite{Bondi:1988fp,Bondi:1989nq} that $O_n$-type divergences do appear at $n$-loop. However, the $O_1$ and $O_2$-type divergences are absent even at 4-loop \cite{Gracey:2016mio}, which implies that they may be absent to all loops. It is also reasonable to conjecture that some other $O_n$ operators are not needed for the renormalizability of Gross-Neveu model either. In the following subsections we will prove that, one can define a parity operator $\chalf$ so that,
\begin{equation}
\chalf O_n=\pm O_n~,
\end{equation}
and only $\chalf$-even operators are required for the renormalizability.

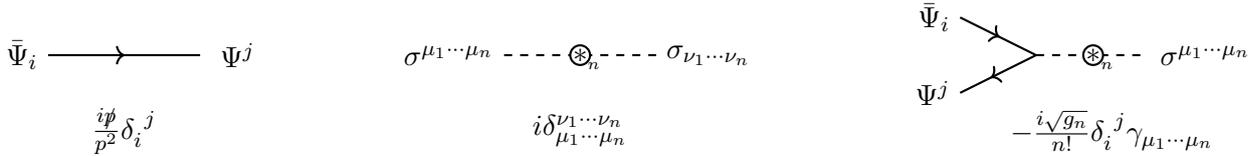
\begin{figure}
  \centering
  \begin{tikzpicture}
    \draw [thick] (-2,0)--(0,0);
    \draw [thick,->] (-1,0)--(-0.99,0);
    \node [left] at (-2,0) {{ $\bar{\Psi}_i$}};
    \node [right] at (0,0) {{ $\Psi^j$}};
    \node [] at (-1,-1) {$\frac{i\slashed{p}}{p^2}\delta^{~j}_{i}$};
    \draw [thick, dashed] (4,0)--(6,0);
    \draw [thick,fill=white] (5,0) circle [radius=0.13];
    \node [] at (5.1,-0.05) {{$\ast_{_n}$}};
    \node [left] at (4,0) {$\sigma^{\mu_1\cdots\mu_n}$};
    \node [right] at (6,0) {$\sigma_{\nu_1\cdots \nu_n}$};
    \node [] at (5,-1) {$i\delta^{\nu_1\cdots\nu_n}_{\mu_1\cdots\mu_n}$};
    \draw [thick] (10,0.5)--(11,0)--(10,-0.5);
    \draw [thick,->] (10.5,0.25)--(10.52,0.24);
    \draw [thick,->] (10.42,-0.29)--(10.4,-0.3);
    \draw [thick,dashed] (11,0)--(12.4,0);
    \draw [thick,fill=white] (11.75,0) circle [radius=0.13];
    \node [] at (11.85,-0.05) {$\ast_{_n}$};
    \node [left] at (10,0.5) {$\bar{\Psi}_i$};
    \node [left] at (10,-0.5) {$\Psi^j$};
    \node [right] at (12.5,0) {$\sigma^{\mu_1\cdots\mu_n}$};
    \node [] at (12,-1) {$-\frac{i\sqrt{g_n}}{n!}\delta^{~j}_{i}\gamma_{\mu_1\cdots\mu_n}$};
   \end{tikzpicture}
  \caption{The Feynman rules of Lagrangian with fermion fields and auxiliary tensor fields. The solid line represents fermion field, with an arrow indicating the direction of fermion flow. The dashed line represents auxiliary tensor field, and the circled astroid with subscript $\circledast_n$ over the dashed line indicates a rank $n$ tensor field $\sigma^{\mu_1\cdots\mu_n}$. This vertex is effectively half of the four-fermion vertex in the original Gross-Neveu Lagrangian, which means that a four-vertex corresponds to two cubic vertices.}\label{fig:FeynmanRule}
\end{figure}

Unlike QED and QCD, in which the fermion chains and fermion loops never intersect, in Gross-Neveu model they intersect at each four-fermion vertex. In order to make the structure of fermion chains more clear, one can introduce a list of auxiliary tensor fields $\sigma_{\mu_1\cdots\mu_n}$, which turns the four-fermion vertices into Yukawa-like interactions,
\begin{equation}
\label{eqn:cubicLagrangian}
\mathcal{L}_{\tiny\mbox{GN}}^{\tiny\mbox{aux}}=\bar{\Psi}_i(i\slashed{\partial}-M)\Psi^i
+\sum_{n=0}^{\infty}\frac{1}{n!}\left(\frac{1}{2}\sigma_{\mu_1\cdots\mu_n}\sigma^{\mu_1\cdots\mu_n}
-\sqrt{g_n}
\sigma_{\mu_1\cdots\mu_n}\bar{\Psi}_i\gamma^{\mu_1\cdots\mu_n}\Psi^i\right)~.
\end{equation}
and the original Lagrangian can be retrieved after integrating out all the auxiliary fields. Since the beta functions are independent of mass, in this paper we will set $M\rightarrow 0$. The Feynman rules\footnote{We take the convention that, in a Feynman diagram, the fermion flows from $\bar{\Psi}$ to $\Psi$, the momenta flows the same direction as the fermion chain, and the gamma matrices are arranged along the opposite direction of the fermion chain. We also have ${\Psi}_i\to \bar{u}_i$ and  $\bar{\Psi}_i\to u_i$.} for $\mathcal{L}_{\tiny\mbox{GN}}^{\tiny\mbox{aux}}$ are provided in Fig.\eqref{fig:FeynmanRule}.
The four-fermion vertex in $\mathcal{L}_{\tiny\mbox{GN}}$ corresponds to a pair of cubic-vertices connected by an auxiliary field in $\mathcal{L}_{\tiny\mbox{GN}}^{\tiny\mbox{aux}}$.

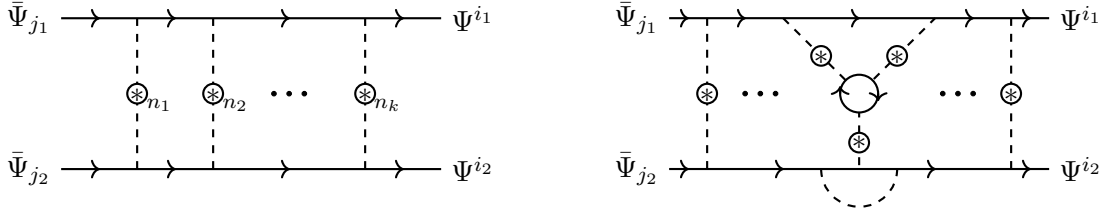
\begin{figure}
  \centering
  \begin{tikzpicture}
    \node [] at (2.5,0) {
    \begin{tikzpicture}
      \draw [thick] (0,-1)--(5,-1) (0,1)--(5,1);
      \draw [thick,->] (0.5,1)--(0.51,1);
      \draw [thick,->] (1.5,1)--(1.51,1);
      \draw [thick,->] (3,1)--(3.01,1);
      \draw [thick,->] (4.5,1)--(4.51,1);
      \draw [thick,->] (0.5,-1)--(0.51,-1);
      \draw [thick,->] (1.5,-1)--(1.51,-1);
      \draw [thick,->] (3,-1)--(3.01,-1);
      \draw [thick,->] (4.5,-1)--(4.51,-1);
      \draw [thick, dashed] (2,-1)--(2,1) (4,-1)--(4,1) (1,-1)--(1,1);
      \draw [fill=white,thick] (1,0) circle [radius=0.13] (2,0) circle [radius=0.13] (4,0) circle [radius=0.13];
      \node [] at (1,-0.02) {$\ast$};
      \node [] at (1.3,-0.15) {$_{n_1}$};
      \node [] at (2.3,-0.15) {$_{n_2}$};
      \node [] at (4.3,-0.15) {$_{n_{k}}$};
      \node [] at (2,-0.02) {$\ast$};
      \node [] at (4,-0.02) {$\ast$};
      \draw [fill=black] (3,0) circle [radius=0.03] (3.2,0) circle [radius=0.03] (2.8,0) circle [radius=0.03];
      \node [left] at (0,1) {$\bar{\Psi}_{j_1}$};
      \node [right] at (5,1) {$\Psi^{i_1}$};
      \node [left] at (0,-1) {$\bar{\Psi}_{j_2}$};
      \node [right] at (5,-1) {$\Psi^{i_2}$};
    \end{tikzpicture}
    };
    \node [] at (10.5,-0.085) {
    \begin{tikzpicture}
      \draw [thick] (0,-1)--(5,-1) (0,1)--(5,1);
      \draw [thick,->] (0.25,1)--(0.26,1);
      \draw [thick,->] (1,1)--(1.01,1);
      \draw [thick,->] (2.5,1)--(2.51,1);
      \draw [thick,->] (4,1)--(4.01,1);
      \draw [thick,->] (4.75,1)--(4.76,1);
      \draw [thick,->] (0.25,-1)--(0.26,-1);
      \draw [thick,->] (1.5,-1)--(1.51,-1);
      \draw [thick,->] (3.5,-1)--(3.51,-1);
      \draw [thick,->] (4.75,-1)--(4.76,-1);
      \draw [thick, dashed] (0.5,-1)--(0.5,1) (4.5,-1)--(4.5,1) (1.5,1)--(2.5,0)--(3.5,1) (2.5,0)--(2.5,-1);
      \draw [fill=white, thick] (2.5,0) circle [radius=0.25];
      \draw [thick,dashed] (2,-1) to [out=270,in=180] (2.5,-1.5) to [out=0,in=270] (3,-1);
      \draw [thick,->] (2.25,0)--(2.25,0.1);
      \draw [thick,->] (2.75,0)--(2.75,-0.1);
      \draw [fill=white,thick] (2,0.5) circle [radius=0.13] (3,0.5) circle [radius=0.13] (2.5,-0.65) circle [radius=0.13] (0.5,0) circle [radius=0.13] (4.5,0) circle [radius=0.13];
      \node [] at (2,0.48) {$\ast$};
      \node [] at (3,0.48) {$\ast$};
      \node [] at (2.5,-0.65) {$\ast$};
      \node [] at (0.5,-0.02) {$\ast$};
      \node [] at (4.5,-0.02) {$\ast$};
      \draw [fill=black] (1,0) circle [radius=0.03] (1.2,0) circle [radius=0.03] (1.4,0) circle [radius=0.03] (4,0) circle [radius=0.03] (3.8,0) circle [radius=0.03] (3.6,0) circle [radius=0.03];
      \node [left] at (0,1) {$\bar{\Psi}_{j_1}$};
      \node [right] at (5,1) {$\Psi^{i_1}$};
      \node [left] at (0,-1) {$\bar{\Psi}_{j_2}$};
      \node [right] at (5,-1) {$\Psi^{i_2}$};
    \end{tikzpicture}
    };
   \end{tikzpicture}
  \caption{Representation of the two types of four-fermion Feynman diagrams. The first type represents diagrams with only two fermion chains connected by interchanging auxiliary tensor fields. The second type represents diagrams with one or more fermion loops connected to the two fermion chains by interchanging tensor fields, and (or) some auxiliary fields are connected to the same fermion chain.}\label{fig:4fermionDiagram}
\end{figure}

The Feynman diagrams of $\mathcal{L}_{\tiny\mbox{GN}}$ can be mapped to the diagrams of $\mathcal{L}_{\tiny\mbox{GN}}^{\tiny\mbox{aux}}$ according to the fermion chains. Taking 1-loop four-fermion Feynman diagrams as example, there are in total two independent topologies. If we use different colors to emphasize the fermion chains, the mapping can be picturised as follows,
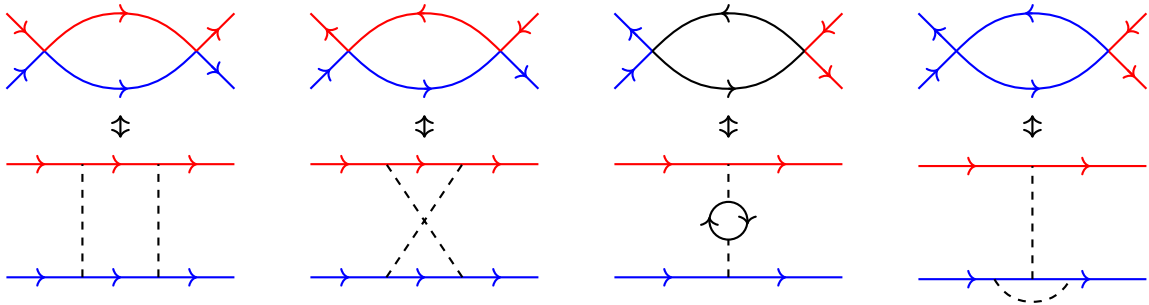
\begin{center}
\begin{tikzpicture}
  \node [] at (1.5,3) {
  \begin{tikzpicture}
    \draw [thick,blue] (0,-0.5)--(0.5,0); 
    \draw [thick,->,blue] (0.25,-0.25)--(0.26,-0.24);
    \draw [thick,red] (0,0.5)--(0.5,0); 
    \draw [thick,->,red] (0.25,0.25)--(0.26,0.24);
    \draw [thick,blue] (2.5,0)--(3,-0.5); 
    \draw [thick,->,blue] (2.7,-0.2)--(2.8,-0.3);
    \draw [thick,red] (2.5,0)--(3,0.5); 
    \draw [thick,->,red] (2.7,0.2)--(2.8,0.3);
    \draw [thick,blue] (0.5,0) to [out=315,in=180] (1.5,-0.5) to [out=0,in=225] (2.5,0); 
    \draw [thick,->,blue] (1.5,-0.5)--(1.6,-0.5);
    \draw [thick,red] (0.5,0) to [out=45,in=180] (1.5,0.5) to [out=0,in=135] (2.5,0);  
    \draw [thick,->,red] (1.5,0.5)--(1.6,0.5);
  \end{tikzpicture}
  };
  \node [] at (1.5,0.75) {
  \begin{tikzpicture}
    \draw [thick,blue] (0,0)--(3,0);
    \draw [thick,->,blue] (0.5,0)--(0.51,0);
    \draw [thick,->,blue] (1.5,0)--(1.51,0);
    \draw [thick,->,blue] (2.5,0)--(2.51,0);
    \draw [thick,red] (0,1.5)--(3,1.5);
    \draw [thick,->,red] (0.5,1.5)--(0.51,1.5);
    \draw [thick,->,red] (1.5,1.5)--(1.51,1.5);
    \draw [thick,->,red] (2.5,1.5)--(2.51,1.5);
    \draw [thick,dashed] (1,0)--(1,1.5) (2,0)--(2,1.5);
  \end{tikzpicture}
  };
  \node [] at (9.5,3) {
  \begin{tikzpicture}
    \draw [thick,blue] (0,-0.5)--(0.5,0); 
    \draw [thick,->,blue] (0.25,-0.25)--(0.26,-0.24);
    \draw [thick,blue] (0,0.5)--(0.5,0); 
    \draw [thick,->,blue] (0.3,0.2)--(0.2,0.3);
    \draw [thick,red] (2.5,0)--(3,-0.5); 
    \draw [thick,->,red] (2.75,-0.25)--(2.85,-0.35);
    \draw [thick,red] (2.5,0)--(3,0.5); 
    \draw [thick,->,red] (2.75,0.25)--(2.74,0.24);
    \draw [thick] (0.5,0) to [out=315,in=180] (1.5,-0.5) to [out=0,in=225] (2.5,0); 
    \draw [thick,->] (1.5,-0.5)--(1.6,-0.5);
    \draw [thick] (0.5,0) to [out=45,in=180] (1.5,0.5) to [out=0,in=135] (2.5,0); 
    \draw [thick,->] (1.5,0.5)--(1.4,0.5);
  \end{tikzpicture}
  };
  \node [] at (9.5,0.75) {
  \begin{tikzpicture}
    \draw [thick,blue] (0,0)--(3,0);
    \draw [thick,->,blue] (0.75,0)--(0.76,0);
    \draw [thick,->,blue] (2.25,0)--(2.26,0);
    \draw [thick,red] (0,1.5)--(3,1.5);
    \draw [thick,->,red] (0.75,1.5)--(0.76,1.5);
    \draw [thick,->,red] (2.25,1.5)--(2.26,1.5);
    \draw [thick,dashed] (1.5,0)--(1.5,1.5);
    \draw [thick,fill=white] (1.5,0.75) circle [radius=0.25];
    \draw [thick,->] (1.25,0.79)--(1.25,0.8);
    \draw [thick,->] (1.75,0.71)--(1.75,0.7);
  \end{tikzpicture}
  };
  \node [] at (5.5,3) {
  \begin{tikzpicture}
    \draw [thick,blue] (0,-0.5)--(0.5,0); 
    \draw [thick,->,blue] (0.25,-0.25)--(0.26,-0.24);
    \draw [thick,red] (0,0.5)--(0.5,0); 
    \draw [thick,->,red] (0.3,0.2)--(0.2,0.3);
    \draw [thick,blue] (2.5,0)--(3,-0.5); 
    \draw [thick,->,blue] (2.75,-0.25)--(2.85,-0.35);
    \draw [thick,red] (2.5,0)--(3,0.5); 
    \draw [thick,->,red] (2.75,0.25)--(2.74,0.24);
    \draw [thick,blue] (0.5,0) to [out=315,in=180] (1.5,-0.5) to [out=0,in=225] (2.5,0); 
    \draw [thick,->,blue] (1.5,-0.5)--(1.6,-0.5);
    \draw [thick,red] (0.5,0) to [out=45,in=180] (1.5,0.5) to [out=0,in=135] (2.5,0); 
    \draw [thick,->,red] (1.5,0.5)--(1.4,0.5);
  \end{tikzpicture}
  };
  \node [] at (5.5,0.75) {
  \begin{tikzpicture}
    \draw [thick,blue] (0,0)--(3,0);
    \draw [thick,->,blue] (0.5,0)--(0.51,0);
    \draw [thick,->,blue] (1.5,0)--(1.51,0);
    \draw [thick,->,blue] (2.5,0)--(2.51,0);
    \draw [thick,red] (0,1.5)--(3,1.5);
    \draw [thick,->,red] (0.5,1.5)--(0.51,1.5);
    \draw [thick,->,red] (1.5,1.5)--(1.51,1.5);
    \draw [thick,->,red] (2.5,1.5)--(2.51,1.5);
    \draw [thick,dashed] (1,0)--(2,1.5) (2,0)--(1,1.5);
  \end{tikzpicture}
  };
  \node [] at (13.5,3) {
  \begin{tikzpicture}
    \draw [thick,blue] (0,-0.5)--(0.5,0); 
    \draw [thick,->,blue] (0.25,-0.25)--(0.26,-0.24);
    \draw [thick,blue] (0,0.5)--(0.5,0); 
    \draw [thick,->,blue] (0.3,0.2)--(0.2,0.3);
    \draw [thick,red] (2.5,0)--(3,-0.5); 
    \draw [thick,->,red] (2.75,-0.25)--(2.85,-0.35);
    \draw [thick,red] (2.5,0)--(3,0.5); 
    \draw [thick,->,red] (2.75,0.25)--(2.74,0.24);
    \draw [thick,blue] (0.5,0) to [out=315,in=180] (1.5,-0.5) to [out=0,in=225] (2.5,0); 
    \draw [thick,->,blue] (1.5,-0.5)--(1.6,-0.5);
    \draw [thick,blue] (0.5,0) to [out=45,in=180] (1.5,0.5) to [out=0,in=135] (2.5,0); 
    \draw [thick,->,blue] (1.5,0.5)--(1.4,0.5);
  \end{tikzpicture}
  };
  \node [] at (13.5,0.63) {
  \begin{tikzpicture}
    \draw [thick,blue] (0,0)--(3,0);
    \draw [thick,->,blue] (0.75,0)--(0.76,0);
    \draw [thick,->,blue] (2.25,0)--(2.26,0);
    \draw [thick,red] (0,1.5)--(3,1.5);
    \draw [thick,->,red] (0.75,1.5)--(0.76,1.5);
    \draw [thick,->,red] (2.25,1.5)--(2.26,1.5);
    \draw [thick,dashed] (1.5,0)--(1.5,1.5);
    \draw [dashed,thick] (1,0) to [out=300,in=180] (1.5,-0.3) to [out=0,in=240] (2,0);
  \end{tikzpicture}
  };
  \draw [thick,<->](1.5,2.15)--(1.5,1.85);
  \draw [thick,<->](5.5,2.15)--(5.5,1.85);
  \draw [thick,<->](9.5,2.15)--(9.5,1.85);
  \draw [thick,<->](13.5,2.15)--(13.5,1.85);
\end{tikzpicture}
\end{center}
Counting the number of gamma matrices in the auxiliary field diagrams, we can confirm that the first and second Feynman diagrams contain four gamma matrices and they indeed contribute to $O_1$-type divergences. While in the third diagram the gamma matrices are contracted in the fermion loop, and in the last diagram the two gamma matrices in the same fermion chain are contracted along the auxiliary field, hence they will not contribute to the $O_1$-type divergences. The latter two diagrams are also named non-1PI-like diagrams in \cite{Tracas:1990wc}, where it is argued that they indeed do not contribute to $O_n$-type divergence. The 1-loop example infers the structure of general-loop Feynman diagrams. Let us classify the diagrams into two types, as represented in Fig.\eqref{fig:4fermionDiagram}.
The first type of four-fermion Feynman diagrams contains two non-intersecting fermion chains, each connects a pair of external legs, and the fermion chains are linked through all possible connections of auxiliary fields. This type contributes to our $O_n$-divergence computations. The other type of diagrams contains either at least one fermion loop, or some auxiliary fields being connected to the same fermion chain, or both. This type will not be considered in the UV divergence computation. By shrinking the auxiliary fields we produce diagrams with four-fermion vertices. We shall emphasize that, in the auxiliary field formulation, the fermion loops never intersect with fermion chains, and they must be connected by the auxiliary fields.

\subsection{The renormalizability of the evanescent operators}
\label{subsec:renormalizability}

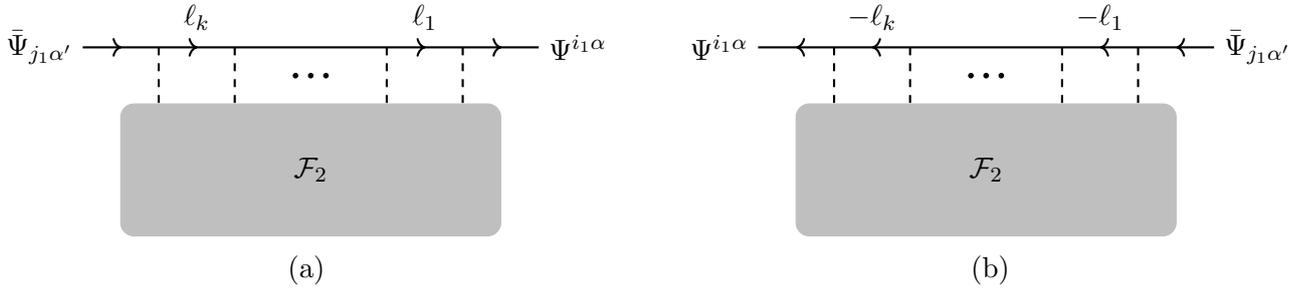
\begin{figure}
  \centering
  \begin{tikzpicture}
    \node [] at (3,-1.75) {
    \begin{tikzpicture}
    \draw [thick] (0,0)--(6,0);
    \draw [thick, dashed] (1,0)--(1,-2) (2,0)--(2,-2) (4,0)--(4,-2) (5,0)--(5,-2);
    \draw [lightgray, fill=lightgray, rounded corners=5] (0.5,-2.5) rectangle (5.5,-0.75);
    \draw [thick,->] (0.5,0)--(0.51,0);
    \draw [thick,->] (1.5,0)--(1.51,0);
    \draw [thick,->] (4.5,0)--(4.51,0);
    \draw [thick,->] (5.5,0)--(5.51,0);
    \draw [fill=black] (3,-0.375) circle [radius=0.03] (3.2,-0.375) circle [radius=0.03] (2.8,-0.375) circle [radius=0.03];
    \node [left] at (0,0) {$\bar{\Psi}_{j_1\alpha'}$};
    \node [right] at (6,0) {${\Psi}^{i_1\alpha}$};
    \node [] at (3,-1.625) {$\mathcal{F}_2$};
    \node [above] at (1.5,0.1) {$\ell_k$};
    \node [above] at (4.5,0.1) {$\ell_1$};
    \end{tikzpicture}
    };
    \node [] at (12,-1.75) {
    \begin{tikzpicture}
    \draw [thick] (0,0)--(6,0);
    \draw [thick, dashed] (1,0)--(1,-2) (2,0)--(2,-2) (4,0)--(4,-2) (5,0)--(5,-2);
    \draw [lightgray, fill=lightgray, rounded corners=5] (0.5,-2.5) rectangle (5.5,-0.75);
    \draw [thick,->] (0.51,0)--(0.5,0);
    \draw [thick,->] (1.51,0)--(1.5,0);
    \draw [thick,->] (4.51,0)--(4.5,0);
    \draw [thick,->] (5.51,0)--(5.5,0);
    \draw [fill=black] (3,-0.375) circle [radius=0.03] (3.2,-0.375) circle [radius=0.03] (2.8,-0.375) circle [radius=0.03];
    \node [right] at (6,0) {$\bar{\Psi}_{j_1\alpha'}$};
    \node [left] at (0,0) {${\Psi}^{i_1\alpha}$};
    \node [] at (3,-1.625) {$\mathcal{F}_2$};
    \node [above] at (1.5,0.1) {$-\ell_k$};
    \node [above] at (4.5,0.1) {$-\ell_1$};
    \end{tikzpicture}
    };
    \node [] at (3,-3.8) {(a)};
    \node [] at (12,-3.8) {(b)};
  \end{tikzpicture}
  \caption{Feynman diagrams contributing to the fermion chain and its companion reversed fermion chain. The vertices $-\frac{i\sqrt{g_{n_i}}}{n_i!}\delta_{i_{n_i}}^{~j_{n_i}}\Gamma^{\mu_i}_{n_i}$ with $i=1,\ldots,k+1$ are labeled from rightmost to leftmost, and the direction of momenta follows the fermion flow. All the remaining part of Feynman diagram has been formally packed into $\mathcal{F}_2$. }\label{fig:fermionChain}
\end{figure}

Now let us discuss the renormalization of the four-fermion vertices, starting from examining a fermion chain connecting $\Psi^{i_1\alpha}$ and $\bar{\Psi}_{j_1\alpha'}$, as sketched in Fig.(\ref{fig:fermionChain}.a).
The integrand of the fermion chain can be written as
\begin{equation}
\mathcal{F}_1=\frac{\Bigl(\Gamma_{n_1}^{\mu_1}\slashed{\ell}_1\Gamma_{n_2}^{\mu_2}
\slashed{\ell}_2
\cdots \Gamma_{n_k}^{\mu_k}\slashed{\ell}_k\Gamma_{n_{k+1}}^{\mu_{k+1}}
\Bigr)^{\alpha}_{~~\alpha'}
}{\ell_1^2\ell_2^2\cdots \ell_k^2}~,
\end{equation}
in which $\Gamma_{n}^{\mu}$ is the abbreviation of $\gamma^{\mu_1\cdots \mu_n}$. The integrand of the remaining part of the diagram will be denoted by $\mathcal{F}_2$. By applying gamma matrix algebra, {\sl e.g.}, the commutation and contraction of gamma matrices as,
\begin{equation}
\label{eqn:gammaMatrix}
\gamma^{\mu}\gamma^{\nu}~\rightarrow~2\eta^{\mu\nu}-\gamma^{\nu}\gamma^{\mu}~~~,~~~ \gamma^{\mu}\gamma_{\mu}~\rightarrow~D~,
\end{equation}
eventually we can rewrite the UV divergences of the diagram into the following form,
\begin{equation}\label{eqn:fermionChainUV}
\mathcal{F}_1\mathcal{F}_2
~~\rightarrow ~~ \sum_m c_m(\gamma_{\nu_1\cdots \nu_m})^{\alpha}_{~\alpha'}~.
\end{equation}
Since gamma matrices contract in pairs, the summation index $m$ in \eqref{eqn:fermionChainUV} must satisfy,
\begin{equation}\label{eqn:fermionChainUVsign}
(-1)^{k-m+\sum_{i=1}^{k+1}n_i}=1~.
\end{equation}

Now let us consider another Feynman diagram by interchanging the positions of $\Psi^{i_1\alpha}$ and $\bar{\Psi}_{j_1\alpha'}$ in the diagram and reversing the direction of fermion chain $\mathcal{F}_1$, without altering the other part of diagram, as in Fig.(\ref{fig:fermionChain}.b). This reversed fermion chain produces
\begin{equation}\label{eqn:fermionChainUVreverse}
\overleftarrow{\mathcal{F}}_1=\frac{\Bigl(\Gamma_{n_{k+1}}^{\mu_{k+1}}(-\slashed{\ell}_k)
\Gamma_{n_k}^{\mu_k}\cdots(-\slashed{\ell}_2)
\Gamma_{n_2}^{\mu_2}(-\slashed{\ell}_1)\Gamma_{n_1}^{\mu_1}
\Bigr)^{\alpha}_{~~\alpha'}}{\ell_1^2\ell_2^2\cdots \ell_k^2}~.
\end{equation}
Let us define an operator $\widehat{\mathcal{R}}$, which reverses the ordering of the gamma matrices. Acting $\widehat{\mathcal{R}}$ on $\Gamma_{n_i}^{\mu_i}$ produces an extra sign as,
\begin{equation}
\widehat{\mathcal{R}}~\Gamma_{n_i}^{\mu_i}=(-1)^{\frac{n_i(n_i-1)}{2}}\Gamma_{n_i}^{\mu_i}~.
\end{equation}
Then the reversed fermion chain can be written as,
\begin{equation}
\overleftarrow{\mathcal{F}}_1=(-1)^{k+\sum_{i=1}^{k+1}\frac{n_i(n_i-1)}{2}}~\widehat{\mathcal{R}}~\mathcal{F}_1~.
\end{equation}
%
%
It can be easily verified that the $\widehat{\mathcal{R}}$ operation commutes with the gamma matrix algebra. The reversed diagram and the original diagram are only differed by the $\widehat{\mathcal{R}}$ operator and an overall sign, thus the UV divergence of the reversed diagram can be computed and reduced in the same manner, which leads to,
\begin{eqnarray}
\overleftarrow{\mathcal{F}}_1\mathcal{F}_2~\rightarrow~&&(-1)^{k+\sum_{i=1}^{k+1}\frac{n_i(n_i-1)}{2}}
\sum_m c_m\widehat{\mathcal{R}}(\gamma_{\nu_1\cdots \nu_m})^{\alpha}_{~~\alpha'}
=(-1)^{k+\sum_{i=1}^{k+1}\frac{n_i(n_i-1)}{2}}\sum_m (-1)^{\frac{m(m-1)}{2}}c_m(\gamma_{\nu_1\cdots \nu_m})^{\alpha}_{~~\alpha'}\nonumber\\
&&=\sum_m (-1)^{\frac{m(m+1)}{2}+\sum_{i=1}^{k+1}\frac{n_i(n_i+1)}{2}}c_m
(\gamma_{\nu_1\cdots \nu_m})^{\alpha}_{~~\alpha'}~,
\end{eqnarray}
where in the last line we have used \eqref{eqn:fermionChainUVsign}. By adding contributions of $\mathcal{F}_1$ and $\overleftarrow{\mathcal{F}}_1$, we obtain
\begin{equation}\label{eqn:fermionChainUVall}
\big(\mathcal{F}_1+\overleftarrow{\mathcal{F}}_1\big)\mathcal{F}_2~~\rightarrow~~ \sum_m\left(1+ (-1)^{\frac{m(m+1)}{2}+\sum_{i=1}^{k+1}\frac{n_i(n_i+1)}{2}}\right)c_m
(\gamma_{\nu_1\cdots \nu_m})^{\alpha}_{~~\alpha'}~.
\end{equation}
The $(\gamma_{\nu_1\cdots \nu_m})^{\alpha}_{~~\alpha'}$ term\footnote{The $(\gamma^{\nu_1\cdots \nu_m})^{\beta}_{~~\beta'}$ term which appears together with $(\gamma_{\nu_1\cdots \nu_m})^{\alpha}_{~~\alpha'}$ is included in $c_m$, and it is not affected by the reversion of $\mathcal{F}_1$ fermion chain.} corresponds to a $O_m$-type divergence, which survives in the condition,
\begin{equation}\label{eqn:surviveCondition}
(-1)^{\frac{m(m+1)}{2}}=(-1)^{\sum_{i=1}^{k+1}\frac{n_i(n_i+1)}{2}}\ .
\end{equation}
We observed that each $O_n$ vertex is associated with a factor $(-1)^{\frac{n(n+1)}{2}}$. Hence the $(-1)^{\frac{n_i(n_i+1)}{2}}$ factors in \eqref{eqn:fermionChainUVall} correspond to the $O_{n_i}$ vertices in the original fermion chain, and the $(-1)^{\frac{m(m+1)}{2}}$ factor corresponds to the resulting $O_m$-type divergence.

Let $\mathcal{C}$ be the charge conjugation operator associated with the $U(1)$ sub-group of $U(N)=U(1)\times SU(N)$, we recognize that the factor of $O_m$-type divergence can be produced by acting the $\mathcal{C}$ on one of the two $\bar{\Psi}\Gamma_n^{\mu}\Psi$'s, which defines,
\begin{equation}
\chalf O_n
:= \frac{1}{2(n!)}\big(\bar{\Psi}_i\gamma_{\mu_1\cdots\mu_n}\Psi^i\big)~
\mathcal{C}\big(\bar{\Psi}_j\gamma^{\mu_1\cdots\mu_n}\Psi^j\big)
=(-1)^{\frac{n(n+1)}{2}}O_n~.\label{eqn:chalf}
\end{equation}
%
For any $n$ that constrained by the condition $(n \mod 4)\equiv 0,3$, we have $\chalf O_n=O_n$, and the operator $O_n$ will be called $\chalf$-even. Otherwise the operator will be called $\chalf$-odd. A fermion chain or a fermion loop is $\chalf$-even (odd) if it contains even (odd) number of $\chalf$-odd vertices. Eqn.\eqref{eqn:surviveCondition} states that the $\chalf$-parities of the UV divergences are the same as the $\chalf$-parities of the fermion chains in Feynman diagrams.

Similar discussions can be applied to fermion loops. If we add the contributions of a fermion loop and its reversing companion, the result vanishes if the fermion loop is $\chalf$-odd,
\begin{equation}
\frac{\mbox{Tr}\Bigl(\Gamma_{n_1}^{\mu_1}
\slashed{\ell}_1\Gamma_{n_2}^{\mu_2}\slashed{\ell}_2
\cdots \Gamma_{n_k}^{\mu_k}\slashed{\ell}_k\Bigr)
}{\ell_1^2\ell_2^2\cdots \ell_k^2}+\Big(\mbox{Reversing~Loop}\Big)~~
\propto ~~1+ (-1)^{\sum_{i=1}^{k}\frac{n_i(n_i+1)}{2}}~.
\end{equation}
This means we only need to consider the four-fermion Feynman diagrams with $\chalf$-even fermion loops. This also guarantees that the two fermion chains in such diagrams must have the same $\chalf$-parity.

In summary, when considering the UV divergences of four-fermion Feynman diagrams, the diagrams with $\chalf$-odd fermion loops can be ignored, and if both fermion chains are $\chalf$-even (odd), the resulting UV divergences are also $\chalf$-even (odd). Thus we propose the $\chalf$-symmetry for Gross-Neveu model,
\begin{quote}
  {\bf The $\chalf$-symmetry}: In Gross-Neveu model, the $\chalf$-odd UV divergences will never appear if all vertices in the Feynman diagrams are $\chalf$-even. Therefore the following $\chalf$-even sector is renormalizable,
\begin{equation}
\mathcal{L}=\bar{\Psi}_i(i\slashed{\partial}-M)\Psi^i
-\frac{1}{2}\sum_{\chalf{\scriptsize\mbox{-even}}}\frac{g_n}{n!}
(\bar{\Psi}_i\gamma_{\mu_1\cdots\mu_n}\Psi^i)(\bar{\Psi}_i\gamma^{\mu_1\cdots\mu_n}\Psi^i)~.\label{eqn:statement}
\end{equation}
\end{quote}
This explains why the previous computations found $O_3, O_4$-type divergences, but not $O_1, O_2$-type ones. We have verified that the $O_5$-type UV divergences do not appear at 5-loop directly using Feynman diagrams. The next $\chalf$-even operator is $O_7$, which first appears at 7-loop. We will compute the contribution of $O_7$ at 7-loop in the next section.

As a side remark, the $\chalf$-parity can also be helpful in the study of fermion operators in other models. For example, in the Gross-Neveu-Yukawa model with Lagrangian,
\begin{equation}
\mathcal{L}_{\tiny\mbox{GNY}}=\bar{\Psi}_i(i\slashed{\partial}-m)\Psi^i
+\frac{1}{2}\partial_{\mu}\phi\partial^{\mu}\phi
-g\phi\bar{\Psi}_i\Psi^i-\frac{\lambda}{4!}\phi^4~,
\end{equation}
the four-fermion operators can also be classified by their $\chalf$-parities.
Since Yukawa coupling is a $\chalf$-even interaction, the $\chalf$-even operators never mix up with the $\chalf$-odd operators. This structure was found in \cite{Ji:2018yaf} by explicit 1-loop computations. By contrast, the gauge interactions are $\chalf$-odd, therefore the $\chalf$-even and the $\chalf$-odd operators do mix up in QED and QCD.

\section{Renormalization of the evanescent operators at 7-loops}
\label{sec:7loop}

As mentioned in the previous section, eqn.(\ref{eqn:chalf}) says that the $O_n$ vertices satisfying $(n\mod 4)\equiv 0,3$ are $\chalf$-even. This includes $O_0,O_3,O_4,O_7,O_8$, {\sl etc}. In order to compute $Z_ng_n$ at $n$-loop, we only need to consider diagrams with the $\chalf$-even vertices $O_{n_1},\cdots, O_{n_V}$ satisfying,
\begin{equation}
\sum_{i=1}^V(n_i+1)=n+1~.
\end{equation}
When $n=7$, the contributing Feynman diagrams for $O_7$-type divergences have the following five types of vertex configuration,
\begin{equation}
\{ n_i \}=(0,0,0,0,0,0,0,0)~~,~~(0,0,0,0,3)~~,~~(0,0,0,4)~~,~~(3,3)~~,~~(7)~.
\end{equation}
The first type of vertex configuration contains eight $O_0$ vertex, and it is a 7-loop contribution, while the other types get contributions from lower loops. In this section, we will introduce techniques to compute these Feynman diagrams, and evaluate the 7-loop beta function of the $O_7$ operator directly.

\subsection{Generating Feynman diagrams from permutation groups}
\label{subsec:7loopDiagram}

Let us first deal with the most complicated Feynman diagrams, {\sl i.e.}, the 7-loop diagram with eight $O_0$ vertices. Most techniques which are useful at computing relatively lower loop Feynman integrals, for example the PV reduction \cite{Passarino:1978jh} and the IBP reduction \cite{Smirnov:2008iw,Maierhofer:2017gsa,Wu:2023upw}, are not efficient at 7-loop. Even the simple task like generating Feynman diagrams can become very challenging with traditional methods. Fortunately, we find that there is a mapping from the permutation group $S_{n+1}$ to $n$-loop Feynman diagrams, which will help us to generate the diagrams and construct their integrands.

To begin with, one notices that not all of the 7-loop Feynman diagrams will produce $O_7$-type divergences. As discussed in \S \ref{subsec:auxiliaryField}, each $n$-loop Feynman diagram contains two fermion chains and some fermion loops, and there are in total $2n$ gamma matrices in the numerator. In order to produce $O_n$-type UV divergences, each fermion chain must have $n$ (anti-symmetrized) gamma matrices.
This rules out any diagrams with fermion loops, in which cases some gamma matrices have to be contracted. Therefore, the contributing Feynman diagrams can only contain two fermion chains connected by $(n+1)$ auxiliary fields, without any fermion loops inside. For those diagrams with some auxiliary fields connecting to the same fermion chain, supposing the two ends of an auxiliary field with momentum $x$ are connected to the same fermion chain, then the fermion chain vanishes after anti-symmetrization,
\begin{equation}
^[\cdots \slashed{\ell}_i(\slashed{\ell}_i+\slashed{x})\cdots \slashed{\ell}_j(\slashed{\ell}_j-\slashed{x})\cdots ^]=0~,
\end{equation}
where the superscript square bracket means all gamma matrices inside are anti-symmetrized. Thus we also rule out the diagrams with auxiliary fields connected to the same fermion chain. The remaining contributing diagrams should have the structure as picturised in the first graph of Fig.(\ref{fig:4fermionDiagram}).

For the contributing Feynman diagrams, we shall consider two different configurations of connecting fermion chains between external legs. Denoting the external fields as $\Psi^{i_1\alpha_1}(p_1)$, $\bar{\Psi}_{i_2\alpha_2}(p_2)$, $\Psi^{i_3\alpha_3}(p_3)$ and $\bar{\Psi}_{i_4\alpha_4}(p_4)$, in the first configuration, $\Psi^{i_1\alpha_1}$ and $\bar{\Psi}_{i_2\alpha_2}$ are connected by one fermion chain, and $\Psi^{i_3\alpha_3}$ and $\bar{\Psi}_{i_4\alpha_4}$ are connected by the other. We will call them $\Psi_1\leftarrow \bar{\Psi}_2$ and $\Psi_3\leftarrow \bar{\Psi}_4$ fermion chains respectively. The second configuration contains $\Psi_1\leftarrow \bar{\Psi}_4$ and $\Psi_3\leftarrow \bar{\Psi}_2$ fermion chains. We will mainly discuss the first configuration, and the UV divergence of the second configuration can be obtained from the first by interchanging $(2\leftrightarrow 4)$.

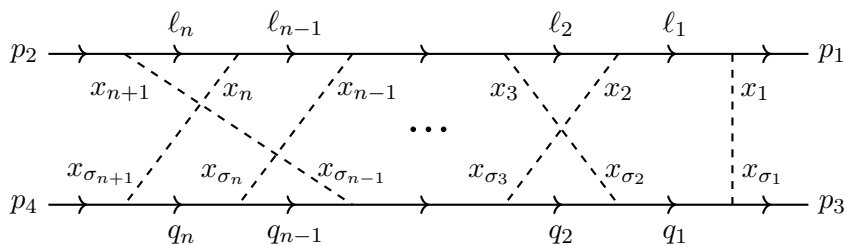
\begin{figure}
\centering
    \begin{tikzpicture}
      \draw [thick] (0,-1)--(10,-1) (0,1)--(10,1);
      \draw [thick,->] (0.5,1)--(0.51,1);
      \draw [thick,->] (1.75,1)--(1.76,1);
      \draw [thick,->] (3.25,1)--(3.251,1);
      \draw [thick,->] (5,1)--(5.01,1);
      \draw [thick,->] (6.75,1)--(6.76,1);
      \draw [thick,->] (8.25,1)--(8.26,1);
      \draw [thick,->] (9.5,1)--(9.51,1);
      \draw [thick,->] (0.5,-1)--(0.51,-1);
      \draw [thick,->] (1.75,-1)--(1.76,-1);
      \draw [thick,->] (3.25,-1)--(3.251,-1);
      \draw [thick,->] (5,-1)--(5.01,-1);
      \draw [thick,->] (6.75,-1)--(6.76,-1);
      \draw [thick,->] (8.25,-1)--(8.26,-1);
      \draw [thick,->] (9.5,-1)--(9.51,-1);
      \draw [thick,dashed] (1,1)--(4,-1) (2.5,1)--(1,-1) (4,1)--(2.5,-1) (6,1)--(7.5,-1) (7.5,1)--(6,-1) (9,1)--(9,-1);
      \node [left] at (0,1) {$p_2$};
      \node [left] at (0,-1) {$p_4$};
      \node [right] at (10,1) {$p_1$};
      \node [right] at (10,-1) {$p_3$};
      \node [] at (1.75,1.4) {$\ell_{n}$};
      \node [] at (3.25,1.4) {$\ell_{n-1}$};
      \node [] at (6.75,1.4) {$\ell_{2}$};
      \node [] at (8.25,1.4) {$\ell_{1}$};
      \node [] at (1.75,-1.4) {$q_{n}$};
      \node [] at (3.25,-1.4) {$q_{n-1}$};
      \node [] at (6.75,-1.4) {$q_{2}$};
      \node [] at (8.25,-1.4) {$q_{1}$};
      \node [] at (0.95,0.5) {$x_{n+1}$};
      \node [] at (2.5,0.5) {$x_{n}$};
      \node [] at (4.2,0.5) {$x_{n-1}$};
      \node [] at (6,0.5) {$x_{3}$};
      \node [] at (7.5,0.5) {$x_{2}$};
      \node [] at (9.3,0.5) {$x_{1}$};
      \node [] at (0.7,-0.6) {$x_{\sigma_{n+1}}$};
      \node [] at (2.3,-0.6) {$x_{\sigma_{n}}$};
      \node [] at (4,-0.6) {$x_{\sigma_{n-1}}$};
      \node [] at (5.8,-0.6) {$x_{\sigma_3}$};
      \node [] at (7.6,-0.6) {$x_{\sigma_2}$};
      \node [] at (9.4,-0.6) {$x_{\sigma_1}$};
      \draw [fill=black] (5,0) circle [radius=0.03] (5.2,0) circle [radius=0.03] (4.8,0) circle [radius=0.03];
    \end{tikzpicture}
\caption{The momentum configuration of Feynman diagrams corresponding to the permutation $\boldsymbol{\sigma}\in S_{n+1}$. The momenta of fermion follows the direction of fermion flow, and the momenta of tensor fields flow from top to bottom.}\label{fig:momentumLabel}
\end{figure}

Let us then analyze the description of Feynman diagrams and the kinematics. Along the opposite direction of the $\Psi_1\leftarrow \bar{\Psi}_2$ fermion chain, we label the momenta of the incident auxiliary fields as $x_1,\ldots,x_{n+1}$.
Let $(\sigma_1,\sigma_2,\ldots,\sigma_{n+1})$ be a permutation of $(1,2,\ldots,n+1)$, and suppose the other end of the $x_{\sigma_k}$ auxiliary field is connected to the $k$-th vertex along the opposite direction of the $\Psi_3\leftarrow\bar{\Psi}_4$ fermion chain, as sketched in Fig.(\ref{fig:momentumLabel}).
Then the diagram can be represented by permutation $\boldsymbol{\sigma}\in S_{n+1}$,
\begin{equation}
\boldsymbol{\sigma}=\begin{pmatrix}
1 & 2 & \cdots & n+1 \\
\sigma_1 & \sigma_2 & \cdots  & \sigma_{n+1}
\end{pmatrix}~,
\end{equation}
and we have a bijection from the Feynman diagrams to the permutation group $S_{n+1}$.
As an example, we have the following mapping,
\begin{center}
\begin{tikzpicture}
    \node [] at (5.5,0) {
    \begin{tikzpicture}
         \draw [thick] (0,-0.5)--(4,-0.5) (0,0.5)--(4,0.5);
         \draw [thick,->] (1,0.5)--(1.01,0.5);
         \draw [thick,->] (2,0.5)--(2.01,0.5);
         \draw [thick,->] (3,0.5)--(3.01,0.5);
         \draw [thick,->] (1,-0.5)--(1.01,-0.5);
         \draw [thick,->] (2,-0.5)--(2.01,-0.5);
         \draw [thick,->] (3,-0.5)--(3.01,-0.5);
         \draw [thick,dashed] (0.5,0.5)--(0.5,-0.5) (1.5,0.5)--(1.5,-0.5) (2.5,0.5)--(2.5,-0.5) (3.5,0.5)--(3.5,-0.5);
         \node [above] at (0.5,0.5) {4};
         \node [above] at (1.5,0.5) {3};
         \node [above] at (2.5,0.5) {2};
         \node [above] at (3.5,0.5) {1};
         \node [below] at (0.5,-0.5) {4};
         \node [below] at (1.5,-0.5) {3};
         \node [below] at (2.5,-0.5) {2};
         \node [below] at (3.5,-0.5) {1};
    \end{tikzpicture}
    };
    \node [] at (1,0) {$\boldsymbol{\sigma}_1=\left({1~2~3~4\atop 1~2~3~4}\right)$};
    \draw [thick,->] (2.5,0)--(3,0);
    \node [] at (14,0) {
    \begin{tikzpicture}
         \draw [thick] (0,-0.5)--(4,-0.5) (0,0.5)--(4,0.5);
         \draw [thick,->] (1,0.5)--(1.01,0.5);
         \draw [thick,->] (2,0.5)--(2.01,0.5);
         \draw [thick,->] (3,0.5)--(3.01,0.5);
         \draw [thick,->] (1,-0.5)--(1.01,-0.5);
         \draw [thick,->] (2,-0.5)--(2.01,-0.5);
         \draw [thick,->] (3,-0.5)--(3.01,-0.5);
         \draw [thick,dashed] (0.5,0.5)--(1.5,-0.5) (1.5,0.5)--(2.5,-0.5) (2.5,0.5)--(0.5,-0.5) (3.5,0.5)--(3.5,-0.5);
         \node [above] at (0.5,0.5) {4};
         \node [above] at (1.5,0.5) {3};
         \node [above] at (2.5,0.5) {2};
         \node [above] at (3.5,0.5) {1};
         \node [below] at (0.5,-0.5) {2};
         \node [below] at (1.5,-0.5) {4};
         \node [below] at (2.5,-0.5) {3};
         \node [below] at (3.5,-0.5) {1};
    \end{tikzpicture}
    };
    \node [] at (9.5,0) {$\boldsymbol{\sigma}_2=\left({1~2~3~4\atop 1~3~4~2}\right)$};
    \draw [thick,->] (11,0)--(11.5,0);
\end{tikzpicture}
\end{center}
As shown in Fig.(\ref{fig:momentumLabel}), we will use $\ell_i$ and $q_i$ to denote the momenta in the upper and lower fermion chains respectively. Then we get
\begin{equation}
\ell_{i-1}=\ell_i-x_i~~~,~~~q_{i-1}=q_i+x_{\sigma_i}~.\label{eqn:momentumRelation}
\end{equation}
As will be shortly discussed, the external momenta $p_i$'s can be set to zero without altering the UV counterterms. After setting all $p_i\to 0$, we get $\ell_1=x_1$ and $q_1=-x_{\sigma_1}$. Using \eqref{eqn:momentumRelation} we obtain
\begin{equation}
\ell_k=\sum_{i=1}^k x_i~~~,~~~q_k=-\sum _{i=1}^{k}x_{\sigma_i}~.
\end{equation}
Hence the fermion chain $\Psi_1\leftarrow \bar{\Psi}_2$ can be written as,
\begin{equation}
\bigl(~^{[~}\slashed{\ell}_1\cdots \slashed{\ell}_n~^{]}~\bigr)^{\alpha_1}_{~~\alpha_2}
=\big( ~^{[~}\slashed{x}_1\cdots \slashed{x}_{n}~^]~\big)^{\alpha_1}_{~~\alpha_2}~,
\end{equation}
and the fermion chain $\Psi_3\leftarrow \bar{\Psi}_4$ equals to,
\begin{equation}\label{eqn:fermionChain4to3}
\big(~^{[~}\slashed{q}_1\cdots \slashed{q}_n^{~~]}~\big)^{\alpha_3}_{~~\alpha_4}
=(-1)^n\big(~^{[~}\slashed{x}_{\sigma_1}\cdots \slashed{x}_{\sigma_n}^{~~]}~\big)^{\alpha_3}_{~~\alpha_4}
=(-1)^n\mbox{sign}(\boldsymbol{\sigma})\big(~^{[~}\slashed{x}_1\cdots \slashed{x}_n^{~~]}~\big)^{\alpha_3}_{~~\alpha_4}~,
\end{equation}
where $\mbox{sign}(\boldsymbol{\sigma})$ is the signature of permutation $\boldsymbol{\sigma}$. They are combined together to generate the complete numerator of integrand as,
\begin{equation}\label{eqn:fermionChainNumerator}
(-1)^n\mbox{sign}(\boldsymbol{\sigma}) \bigl(~^{[~}\slashed{\ell}_1\cdots \slashed{\ell}_n~^{]}~\bigr)^{\alpha_1}_{~~\alpha_2} \bigl(~^{[~}\slashed{\ell}_1\cdots \slashed{\ell}_n~^{]}~\bigr)^{\alpha_3}_{~~\alpha_4}~.
\end{equation}
This expression is independent of the parametrization of loop momenta. The complete $n$-loop integrand of all contributing Feynman diagrams will be denoted by\footnote{The $(-1)^n$ factor in \eqref{eqn:fermionChainNumerator} is cancelled by the $i$ factors in the fermion propagators.},
\begin{equation}\label{eqn:fermionChainIntegrand}
\mathcal{I}^{(n)}= (-ig)^{n+1}\delta_{i_1}^{i_2}\delta_{i_3}^{i_4} \bigl(~^{[~}\slashed{\ell}_1\cdots \slashed{\ell}_n~^{]}~\bigr)^{\alpha_1}_{~~\alpha_2} \bigl(~^{[~}\slashed{\ell}_1\cdots \slashed{\ell}_n~^{]}~\bigr)^{\alpha_3}_{~~\alpha_4}\sum_{\boldsymbol{\sigma}\in S_{n+1}}\mbox{sign}(\boldsymbol{\sigma})\mathcal{I}_{\boldsymbol{\sigma}}~,
\end{equation}
where $\mathcal{I}_{\boldsymbol{\sigma}}$ is the scalar integral associated with the denominator of the diagram corresponding to the permutation $\boldsymbol{\sigma}$. The integrand (\ref{eqn:fermionChainIntegrand}) includes all the contributions of 7-loop Feynman diagrams with eight $O_0$ vertices for $O_7$-type UV divergence, and what we should do next is to extract the UV information by evaluating the two-dimensional integrals.

\subsection{From two-dimensional tensor integrals to four-dimensional scalar integrals}
\label{subsec:7loopIntegral}

Currently the graphical function method \cite{Schnetz:2013hqa,Borinsky:2021gkd} remains the exclusive approach for calculating 7-loop Feynman integrals, but right now the method only works for spacetime dimension $D\ge 4$ integrals without numerators\footnote{See \cite{Schnetz:2024qqt} for new progresses on tensor integrals.}. So if we want to evaluate the two-dimensional 7-loop integrals with tensor structures of loop momenta in the numerator by graphical function method, the only possibility is if the two-dimensional tensor integrals can be converted to $D\ge 4$-dimensional scalar integrals. Here we want to demonstrate that, if only UV divergences are desired, we can indeed extract equivalent UV divergences from evaluating four-dimensional scalar integrals instead of two-dimensional tensor integrals.

The correlation functions in Gross-Neveu model contain logarithmic (divergence degree $\omega=0$) and quadratic ($\omega=2$) divergences. The logarithmic divergences correspond to four-point sub-graphs, and the quadratic divergences correspond to two-point sub-graphs. The Feynman diagrams, which we are concerned with, never contain two-point sub-graphs. This is because the two-point sub-graphs correspond to the Feynman diagrams of $\langle\Psi \bar{\Psi}\rangle$ correlation function, which will contain one fermion chain and some fermion loops. As has been discussed in \S\ref{subsec:7loopDiagram}, diagrams with fermion loops can be ignored. While in those diagrams without fermion loops, the auxiliary fields can only be connected to the single fermion chain, which will not produce $O_n$-type divergence either. Therefore, the diagrams we considered only contain logarithmic UV divergences.

If a correlation function $\mathcal{G}$ only contains the logarithmic UV divergence, then according to the IR rearrangement technique \cite{Vladimirov:1979zm,Chetyrkin:1980pr,Caswell:1981ek} and following the standard $R^\ast$-operation procedure \cite{Chetyrkin:1982nn,CHETYRKIN1984419,Larin:2002sc}, we have the conclusion that, applying one or the combination of the following operations to the original integrand does not introduce extra UV divergence to the {\sl renormalized correlation function},
\begin{flalign*}
&\begin{array}{l}
  ~~~\diamond~~\mbox{Add~the~same~mass}~m~\mbox{to~all~propagators}. \\
  ~~~\diamond~~\mbox{Set~the~mass}~m\to 0. \\
  ~~~\diamond~~\mbox{In~the~numerators,~set~all~external~momenta}~p_i\to 0.\\
  ~~~\diamond~~\mbox{Set~one~or~more~external~momenta}~p_i\to 0, \mbox{but~without~introducing~infrared~divergences.}\\
  ~~~\diamond~~\mbox{Resume~external~momenta}~p_i~\mbox{which~were~set~to}~0~\mbox{in~other~operations.}
\end{array}&
\end{flalign*}
This is equivalent to claim that, the UV counterterms of the original Feynman diagrams in $\mathcal{G}$ are not altered by these operations.

We can apply above mentioned operations to the tensor integrals (\ref{eqn:fermionChainIntegrand}). Let us generally express the $\mathcal{I}_{\boldsymbol{\sigma}}$ in (\ref{eqn:fermionChainIntegrand}) as a $D$-dimensional integral,
\begin{equation}
\mathcal{I}_{\boldsymbol{\sigma}}^{{\tiny\mbox{dim-{\sl D}}}}=\frac{1}{\ell_1^2\cdots \ell_{n}^2q_1^2\cdots q_{n}^2}~,
\end{equation}
in which $\ell_i$, $q_i$ are the loop momenta of the $\Psi_1\leftarrow\bar{\Psi}_2$, $\Psi_3\leftarrow\bar{\Psi}_4$ fermion chains respectively. We would like to prove that the UV counterterm of (\ref{eqn:fermionChainIntegrand}) is equivalent to some scalar integrals in higher dimension. Let us start from the Feynman integral,
\begin{equation}\label{eqn:dim-shift-1}
\mathcal{F}^{\tiny\mbox{dim-{\sl D}}}=\frac{i^{2n}\big(~^{[~}\slashed{\ell}_1\cdots\slashed{\ell}_n ~^{]}~\big)^{\alpha_1}_{~~ \alpha_2}\big(~
^{[~}\slashed{q}_1\cdots\slashed{q}_n^{~~]}~\big)^{\alpha_3}_{~~\alpha_4}
}{\ell_1^2\cdots \ell_{n}^2q_1^2\cdots q_{n}^2}~.
\end{equation}
Firstly we add the same mass $m$ to all propagators, and then set external momenta $p_i\rightarrow 0$, which rewrites the original integral $\mathcal{F}$ to,
\begin{equation}
\mathcal{F}_{\tiny\mbox{step-1}}=
\left.\frac{\mbox{sign}(\boldsymbol{\sigma})\big(~^{[~}\slashed{\ell}_1\cdots\slashed{\ell}_n~^{]}~\big)^{\alpha_1}_{~~\alpha_2}
\big(~^{[~}\slashed{\ell}_1\cdots\slashed{\ell}_n~^{]}~\big)^{\alpha_3}_{~~\alpha_4}}{(\ell_1^2-m^2)\cdots (\ell_{n}^2-m^2)(q_1^2-m^2)\cdots (q_{n}^2-m^2)}\right|_{p_i=0}~,
\end{equation}
where in the derivation we have used relation \eqref{eqn:fermionChain4to3}. As discussed previously, the UV counterterms of $\mathcal{F}$ and $\mathcal{F}_{\tiny\mbox{step-1}}$ are the same. Next we perform a PV reduction \cite{Passarino:1978jh} to $\mathcal{F}_{\tiny\mbox{step-1}}$ and get,
\begin{equation}
\mathcal{F}_{\tiny\mbox{step-2}}=\left.\frac{(-1)^n}{(-D)^n}
\frac{\mbox{sign}(\boldsymbol{\sigma}){G}(\ell_1,\ldots,\ell_n)
\big(\gamma^{\mu_1\cdots\mu_n}\big)^{\alpha_1}_{~~ \alpha_2}\big(\gamma_{\mu_1\cdots\mu_n}\big)^{\alpha_3}_{~~\alpha_4}
}{(\ell_1^2-m^2)\cdots (\ell_{n}^2-m^2)(q_1^2-m^2)\cdots (q_{n}^2-m^2)}\right|_{p_i=0}~,
\end{equation}
in which $G(\ell_1,\ldots,\ell_n)$ is the Gram determinant. Then we use dimensional shifting formula \cite{Tarasov:1996br,Tarasov:1997kx} to convert the $D$-dimensional integral to a $(D+2)$-dimensional integral without numerators, and get
\begin{equation}
\mathcal{F}_{\tiny\mbox{step-3}}=\left.\frac{(2^{-n})\mbox{sign}(\boldsymbol{\sigma})
\big(\gamma^{\mu_1\cdots\mu_n}\big)^{\alpha_1}_{~~\alpha_2}
\big(\gamma_{\mu_1\cdots\mu_n}\big)^{\alpha_3}_{~~\alpha_4}}{(\ell_1^2-m^2)\cdots (\ell_{n}^2-m^2)(q_1^2-m^2)\cdots (q_{n}^2-m^2)}\right|_{D\rightarrow D+2~,~p_i=0}~.
\end{equation}
Note that both the PV reduction and dimensional shifting produce $D$-dependent factors, which may alter the counterterms. However, these two factors magically cancel, and $\mathcal{F}_{\tiny\mbox{step-3}}$ does not explicitly depend on $D$. This indicates that $\mathcal{F}_{\tiny\mbox{step-1}}$ and $\mathcal{F}_{\tiny\mbox{step-3}}$ have the same UV counterterm. This is indeed the case, and can be proven by closely examining their sub-divergences. We have also directly verified this relation for various multi-loop integrals. Finally, we can retrieve $p_i$ and remove the mass, and obtain $(D+2)$-dimensional integral without changing the UV counterterms,
\begin{equation}
\mathcal{F}^{\tiny\mbox{dim-{\sl (D+2)}}}=\left.\frac{\mbox{sign}(\boldsymbol{\sigma})}{2^n}
\frac{\big(\gamma^{\mu_1\cdots\mu_n}\big)^{\alpha_1}_{~~ \alpha_2}\big(\gamma_{\mu_1\cdots\mu_n}\big)^{\alpha_3}_{~~\alpha_4}}{\ell_1^2\cdots \ell_{n}^2q_1^2\cdots q_{n}^2}\right|_{D\rightarrow D+2}~.
\end{equation}

Applying above discussion to \eqref{eqn:fermionChainIntegrand}, the two-dimensional tensor integral can be eventually transferred to the equivalent four-dimensional scalar integral as
\begin{equation}\label{eqn:4Dintegrand}
\mathcal{I}^{(n)}= \frac{(-ig)^{n+1}}{2^n}\delta_{i_1}^{i_2}\delta_{i_3}^{i_4}
\big(\gamma^{\mu_1\cdots\mu_n}\big)^{\alpha_1}_{~~ \alpha_2}
\big(\gamma_{\mu_1\cdots\mu_n}\big)^{\alpha_3}_{~~\alpha_4}
\sum_{\boldsymbol{\sigma}\in S_{n+1}}\mbox{sign}(\boldsymbol{\sigma})\mathcal{I}_{\boldsymbol{\sigma}}^{\tiny\mbox{dim-4}}~.
\end{equation}
%
%
%
So we have reduced the problem to the evaluation of four-dimensional scalar integrals. The contributing Feynman diagrams form a subset of the $\langle \phi\bar{\phi}\phi\bar{\phi}\rangle$ diagrams in the quartic scalar theory with $U(1)$ symmetry,
\begin{equation}
\mathcal{L}_{\phi^4}=\partial_{\mu}\bar{\phi}\partial^{\mu}\phi-\frac{g}{4}(\bar{\phi}\phi)^2
~~~\mbox{or~with~auxiliary~field}~~~  \mathcal{L}_{\phi^4}^{\tiny\mbox{aux}}=\partial_{\mu}\bar{\phi}\partial^{\mu}\phi+\frac{1}{2}\sigma^2
-\sqrt{\frac{g}{2}}\sigma\bar{\phi}\phi~.
\end{equation}

Now we are confronting the task of computing four-dimensional four-point 7-loop Feynman diagrams. Some of these diagrams can be evaluated directly using graphical function method, with \verb!GraphicalFunction! command in \verb!HyperlogProcedures! package. But the more effective approach requires the aid of large momentum expansion \cite{Smirnov:2002pj}. We will explain this technique in the next subsection. In order to determine the UV counterterms, we merely require the leading order terms in the large momentum expansion, which correspond to the propagator-type integrals without numerators, and can be evaluated with remarkable efficiency via the graphical function method.

\subsection{Computing UV divergence via large momentum expansion}
\label{subsec:7loopOPE}

The graphical function method, currently the sole analytical framework capable of evaluating Feynman integrals up to 7-loops, has been exclusively applied to scalar field theories, {\sl e.g.}, quartic scalar theory \cite{Schnetz:2022nsc}. In this paper we present its first extension to a system with fermionic degrees of freedom.

To remind, what we want to compute is the UV divergences of four-point scalar integrals of quartic scalar theory in four-dimension. In order to use the graphical function method, we first transform the four-point integrals to coordinate space,
\begin{equation}
\mathcal{I}_{\boldsymbol{\sigma}}^{\tiny\mbox{dim-4}}(p_1,p_2,p_3,p_4)~~~\to~~~ \widetilde{\mathcal{I}}_{\boldsymbol{\sigma}}^{\tiny\mbox{dim-4}}(z_1,z_2,z_3,z_4)~,
\end{equation}
where $z_1,z_2,z_3,z_4$ are the positions of $\Psi^{i_1},\bar{\Psi}_{i_2}, \Psi^{i_3},\bar{\Psi}_{i_4}$ respectively in coordinate space. As discussed in \cite{Schnetz:2022nsc,Huang:2024hsn}, when $z_2$ approaches $z_1$ while $z_3, z_4$ are relatively far away, the integral can be expanded into an asymptotic series of variable $z=z_{12}:= z_1-z_2$ as,
\begin{equation}
\widetilde{\mathcal{I}}_{\boldsymbol{\sigma}}^{\tiny\mbox{dim-4}}(z_1,z_2,z_3,z_4)
~~\sim~~ \sum_{n,k,a}C_{n,k,a}^{\mu_1\cdots \mu_a}(z_{13},z_{14})~\big(z^2\big)^n~\big(\ln^k|z|\big)~\big(z_{\mu_1}\cdots z_{\mu_a}\big)~,
\end{equation}
which is the large momentum expansion in coordinate space. In momentum space, this means the two fields $\Psi^{i_1}$ and $\bar{\Psi}_{i_2}$ are taken to the large momentum limit. The expansion terms $(z^2)^n\ln^k|z|$ are independent to each other. As analyzed in \cite{Huang:2024hsn}, the leading order term ($n=0$) of the expansion can be conveniently obtained by summing over contributions of all minimal-cuts, and for quartic scalar theory in four-dimension the minimal number of cuts is 2. By applying 2-cut to the considered Feynman diagram, it will split into two sub-diagrams, one {\sl hard} sub-diagram with all hard external fields and the other {\sl soft} sub-diagram with all soft external fields. Then the Feynman integral can be factorized as a product of {\sl hard integral} $\mathbb{H}_{\boldsymbol{\sigma}\alpha}(z)$ and {\sl soft integral} $\mathbb{S}_{\boldsymbol{\sigma}\alpha}(z_{13},z_{14})$ as,
\begin{equation}
\widetilde{\mathcal{I}}_{\boldsymbol{\sigma}}^{\tiny\mbox{dim-4}}(z_1,z_2,z_3,z_4)
~~\sim~~\sum_{\alpha}\mathbb{H}_{\boldsymbol{\sigma}\alpha}(z)
\mathbb{S}_{\boldsymbol{\sigma}\alpha}(z_{13},z_{14})~.
\end{equation}
The hard integral is a propagator-type integral of the form,
\begin{equation}
\mathbb{H}_{\boldsymbol{\sigma}\alpha}(z)=\mathbf{H}_{\boldsymbol{\sigma}\alpha}(\epsilon)~(z^2)^{(L_{\alpha}+1)\epsilon}
=\mathbf{H}_{\boldsymbol{\sigma}\alpha}(\epsilon)~\Bigl(1+2(L_{\alpha}+1)\epsilon\ln|z|+\cdots\Bigr)~,
\end{equation}
in which $L_{\alpha}$ is the loop order of the hard sub-diagram.

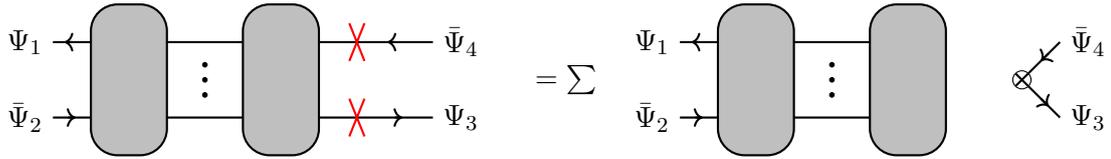
\begin{figure}
\centering
    \begin{tikzpicture}
      \node [] at (2.5,1) {
      \begin{tikzpicture}
        \draw [thick] (0,0.5)--(5,0.5) (0,1.5)--(5,1.5);
        \draw [fill=black] (2,1) circle [radius=0.03] (2,0.8) circle [radius=0.03] (2,1.2) circle [radius=0.03];
        \draw [thick,fill=lightgray,rounded corners=10] (0.5,0) rectangle (1.5,2);
        \draw [thick,fill=lightgray,rounded corners=10] (2.5,0) rectangle (3.5,2);
        \draw [thick,->] (0.2,0.5)--(0.3,0.5);
        \draw [thick,->] (0.25,1.5)--(0.15,1.5);
        \draw [thick,->] (4.5,0.5)--(4.6,0.5);
        \draw [thick,->] (4.5,1.5)--(4.4,1.5);
        \draw [thick,red] (4.1,0.25)--(3.9,0.75) (3.9,0.25)--(4.1,0.75);
        \draw [thick,red] (4.1,1.25)--(3.9,1.75) (3.9,1.25)--(4.1,1.75);
        \node [left] at (0,0.5) {$\bar{\Psi}_2$};
        \node [left] at (0,1.5) {$\Psi_1$};
        \node [right] at (5,0.5) {$\Psi_3$};
        \node [right] at (5,1.5) {$\bar{\Psi}_4$};
      \end{tikzpicture}
      };
      \node [] at (10.75,1) {
      \begin{tikzpicture}
        \draw [thick] (0,0.5)--(3,0.5) (0,1.5)--(3,1.5);
        \draw [fill=black] (2,1) circle [radius=0.03] (2,0.8) circle [radius=0.03] (2,1.2) circle [radius=0.03];
        \draw [thick,fill=lightgray,rounded corners=10] (0.5,0) rectangle (1.5,2);
        \draw [thick,fill=lightgray,rounded corners=10] (2.5,0) rectangle (3.5,2);
        \draw [thick,->] (0.2,0.5)--(0.3,0.5);
        \draw [thick,->] (0.25,1.5)--(0.15,1.5);
        \draw [thick] (4.5,1)--(5,1.5) (4.5,1)--(5,0.5);
        \draw [thick,->] (4.85,1.35)--(4.75,1.25);
        \draw [thick,->] (4.75,0.75)--(4.85,0.65);
        \draw [fill=white] (4.5,1) circle [radius=0.12];
        \draw [thick] (4.42,0.92)--(4.58,1.08) (4.58,0.92)--(4.42,1.08);
        \node [left] at (0,0.5) {$\bar{\Psi}_2$};
        \node [left] at (0,1.5) {$\Psi_1$};
        \node [right] at (5,0.5) {$\Psi_3$};
        \node [right] at (5,1.5) {$\bar{\Psi}_4$};
      \end{tikzpicture}
      };
      \node [] at (6.75,1) {$=\sum $};
    \end{tikzpicture}
\caption{The $\mathcal{O}(\ln|z|)$ contributions of four-fermion integrand under large momentum expansion. The 2-cut on the legs $z_3,z_4$ separates Feynman graph into a hard graph of two-point propagator-type integral with $\Psi_1,\bar{\Psi}_2$ fields and a soft graph of tree-level form factor with $\Psi_3,\bar{\Psi}_4$ fields. The other 2-cuts with such separation only exist in Feynman diagrams not contributing to the evanescent operator computation. }\label{fig:OPEcut}
\end{figure}

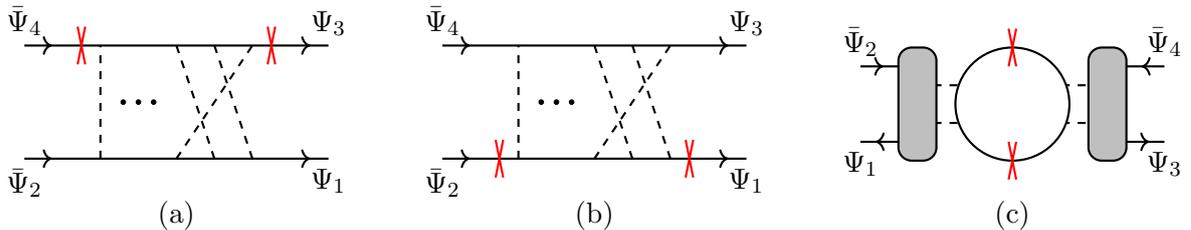
\begin{figure}
\centering
    \begin{tikzpicture}
      \node [] at (2.5,1){
      \begin{tikzpicture}
      \draw [thick] (0,0)--(4,0) (0,1.5)--(4,1.5);
      \draw [thick,dashed] (1,0)--(1,1.5) (2,0)--(3,1.5) (2.5,0)--(2,1.5) (3,0)--(2.5,1.5);
      \draw [fill=black] (1.5,0.75) circle [radius=0.03] (1.3,0.75) circle [radius=0.03] (1.7,0.75) circle [radius=0.03];
      \draw [thick,->] (0.35,0)--(0.36,0);
      \draw [thick,->] (0.35,1.5)--(0.36,1.5);
      \draw [thick,->] (3.75,0)--(3.76,0);
      \draw [thick,->] (3.75,1.5)--(3.76,1.5);
      \node [above] at (0,1.5) {$\bar{\Psi}_4$};
      \node [below] at (0,0) {$\bar{\Psi}_2$};
      \node [above] at (4,1.5) {${\Psi}_3$};
      \node [below] at (4,0) {${\Psi}_1$};
      \draw [thick,red] (0.7,1.25)--(0.8,1.75) (0.8,1.25)--(0.7,1.75);
      \draw [thick,red] (3.2,1.25)--(3.3,1.75) (3.3,1.25)--(3.2,1.75);
      \end{tikzpicture}
      };
      \node [] at (8,1){
      \begin{tikzpicture}
      \draw [thick] (0,0)--(4,0) (0,1.5)--(4,1.5);
      \draw [thick,dashed] (1,0)--(1,1.5) (2,0)--(3,1.5) (2.5,0)--(2,1.5) (3,0)--(2.5,1.5);
      \draw [fill=black] (1.5,0.75) circle [radius=0.03] (1.3,0.75) circle [radius=0.03] (1.7,0.75) circle [radius=0.03];
      \draw [thick,->] (0.35,0)--(0.36,0);
      \draw [thick,->] (0.35,1.5)--(0.36,1.5);
      \draw [thick,->] (3.75,0)--(3.76,0);
      \draw [thick,->] (3.75,1.5)--(3.76,1.5);
      \node [above] at (0,1.5) {$\bar{\Psi}_4$};
      \node [below] at (0,0) {$\bar{\Psi}_2$};
      \node [above] at (4,1.5) {${\Psi}_3$};
      \node [below] at (4,0) {${\Psi}_1$};
      \draw [thick,red] (0.7,-0.25)--(0.8,0.25) (0.8,-0.25)--(0.7,0.25);
      \draw [thick,red] (3.2,-0.25)--(3.3,0.25) (3.3,-0.25)--(3.2,0.25);
      \end{tikzpicture}
      };
      \node [] at (13.5,1){
      \begin{tikzpicture}
      \draw [thick,dashed] (0.75,0.5)--(3.25,0.5) (0.75,1)--(3.25,1);
      \draw [thick] (0,0.25)--(0.75,0.25) (0,1.25)--(0.75,1.25) (3.25,0.25)--(4,0.25) (3.25,1.25)--(4,1.25);
      \draw [thick,fill=lightgray,rounded corners=5] (0.5,0) rectangle (1,1.5);
      \draw [thick,fill=lightgray,rounded corners=5] (3,0) rectangle (3.5,1.5);
      \draw [thick,fill=white] (2,0.75) circle [radius=0.75];
      \draw [thick,->] (0.29,1.25)--(0.3,1.25);
      \draw [thick,->] (0.21,0.25)--(0.2,0.25);
      \draw [thick,->] (3.71,1.25)--(3.7,1.25);
      \draw [thick,->] (3.79,0.25)--(3.8,0.25);
      \node [above] at (0,1.25) {$\bar{\Psi}_2$};
      \node [below] at (0,0.25) {${\Psi}_1$};
      \node [above] at (4,1.25) {$\bar{\Psi}_4$};
      \node [below] at (4,0.25) {${\Psi}_3$};
      \draw [thick,red] (1.95,-0.25)--(2.05,0.25) (2.05,-0.25)--(1.95,0.25);
      \draw [thick,red] (1.95,1.25)--(2.05,1.75) (2.05,1.25)--(1.95,1.75);
      \end{tikzpicture}
      };
      \node [] at (2.5,-0.5) {(a)};
      \node [] at (8,-0.5) {(b)};
      \node [] at (13.5,-0.5) {(c)};
    \end{tikzpicture}
\caption{In the auxiliary field formulation, the three possible 2-cut configurations that separate hard fields $\bar{\Psi}_2, \Psi_1$ and soft fields $\bar{\Psi}_4,\Psi_3$. The cut can only be applied to the fermion line. (a) The 2-cut configuration contributing to $\mathcal{O}(\ln|z|)$. (b) The 2-cut configuration contributing to $\mathcal{O}(\ln^0|z|)$. (c) The 2-cut configuration of Feynman diagrams with fermion loops.}\label{fig:OPEcutAuxi}
\end{figure}

As concluded in \cite{Schnetz:2022nsc}, both $\mathcal{O}(\ln^0|z|)$ and $\mathcal{O}(\ln|z|)$ terms are adequate to fix the UV counterterms, but the evaluation of $\mathcal{O}(\ln|z|)$ terms are relatively simpler. For the set of $n$-loop diagrams in consideration, the only cut configurations contributing to $\mathcal{O}(\ln |z|)$ terms are those in which only the two external legs $z_3$ and $z_4$ are cut, which split the diagram into a hard diagram of loop-level two-point propagator-type integral and a soft diagram of tree-level form factor, as shown in Fig.(\ref{fig:OPEcut}).
To see this, let us go to the auxiliary field formulation. Each four-fermion vertex has been represented by two cubic vertices connected by an auxiliary field, and we should consider all possible 2-cuts that separate the soft fields $\Psi^{i_3},\bar{\Psi}_{i_4}$ from hard fields $\Psi^{i_1},\bar{\Psi}_{i_2}$. The cuts must be restricted to the fermion lines, as the auxiliary field lines lack physical propagators. The Feynman diagrams that contributing to the considered UV divergences are those only containing two fermion chains connected by auxiliary fields in between. There are only two possible ways of separating the hard and soft sub-diagrams by 2-cut for these diagrams, as picturised in Fig.(\ref{fig:OPEcutAuxi}.a) and (\ref{fig:OPEcutAuxi}.b). Cuts on other fermion propagators will either separate the two hard fields, or keep the soft and hard fields in the same sub-diagram, which are not allowed by our consideration. The 2-cut shown in Fig.(\ref{fig:OPEcutAuxi}.a) is exactly the one we have illustrated in Fig.(\ref{fig:OPEcut}), while the 2-cut shown in Fig.(\ref{fig:OPEcutAuxi}.b) separate the diagram into a loop-level soft sub-diagram of two-point form factor and two disconnected points of hard fields\footnote{This disconnected diagram produces $\mathcal{O}(\ln^0 |z|)$ terms but not $\mathcal{O}(\ln |z|)$ terms. So by considering $\mathcal{O}(\ln |z|)$ terms, one circumvents the evaluation of the disconnected diagrams, which can be more difficult than the other diagrams.}, which dose not depend on $z$. There is another possibility of separating the hard and soft part by 2-cut, which requires applying 2-cut on the propagators of certain fermion loop, as shown in Fig.(\ref{fig:OPEcutAuxi}.c). However Feynman diagrams with fermion loops will not contribute to the UV divergence of considered evanescent operators, and have already been ruled out before. Hence the 2-cut configuration shown in Fig.(\ref{fig:OPEcut}) is sufficient for our UV divergences computation.
%



For this cut-configuration, the soft sub-diagrams are the same for all graphs,
\begin{equation}
\mathbb{S}_{\boldsymbol{\sigma}\alpha}(z_{13},z_{14})
=\frac{1}{(z_{13}^2z_{14}^2)^{\frac{D-2}{2}}}~,
\end{equation}
and they become an unimportant overall factor. We only need to focus on the hard integrals. The hard integrals are always a propagator-type integral without numerators, and can be evaluated to 7-loops using graphical function method \cite{Schnetz:2022nsc} with \verb!TwoPoint! command in \verb!HyperlogProcedures! package. After getting the expression of the hard integrals, the UV divergence can be obtained from the expression. Then we can use the $\overline{\mbox{MS}}$ scheme to compute the anomalous dimensions of the operators.

\subsection{Contributions of more Feynman diagrams with evanescent vertices}
\label{subsec:7loopMore}

In the previous subsections, we focused on the techniques for Feynman diagrams with only $O_0$ vertices. However these techniques can be similarly applied to the diagrams with evanescent vertices. The difference of the latter type Feynman diagrams emerges from the gamma matrices associated with the $O_n$ vertices, otherwise they behave totally in the same manner. For $O_7$-type divergences, we should also consider the vertex configurations of $\{n_i\}=(0,0,0,0,3)$, $(0,0,0,4)$, $(3,3)$ and $(7)$, which will be computed by 4-loop, 3-loop, 1-loop and tree Feynman diagrams respectively. In this subsection, we will again rewrite them as scalar integrals in four-dimension, being appropriate for evaluating via graphical function method.

Let us consider a two-dimensional $(V-1)$-loop diagram with $V$ generic four-fermion vertices. Suppose along the opposite direction of the $\Psi_1\leftarrow\bar{\Psi}_2$ fermion chain, the vertices are $(O_{n_1},\cdots, O_{n_V})$, and along the opposite direction of the $\Psi_3\leftarrow\bar{\Psi}_4$ fermion chain, the vertices are $(O_{n_{\sigma_1}},\cdots, O_{n_{\sigma_V}})$. Then follow a similar kinematics as in Fig.(\ref{fig:momentumLabel}), and define the length-$V$ permutation $\boldsymbol{\sigma}\in S_{V}$, we can write the integrand of Feynman diagram as,
\begin{equation}
\mathcal{F}=i^{2V-2}\prod_{j=1}^V\frac{-ig_{n_j}}{n_j!}\Bigl(
~^[~\Gamma_{n_1}^{\mu_1}\slashed{\ell}_1
\Gamma_{n_2}^{\mu_2}\cdots \slashed{\ell}_{V-1}\Gamma_{n_V}^{\mu_V}~^]~\Bigr)^{\alpha_1}_{~~\alpha_2}\times \Bigl(~^[~\Gamma_{n_{\sigma_1}}^{\mu_{\sigma_1}}\slashed{q}_1
\Gamma_{n_{\sigma_2}}^{\mu_{\sigma_2}}
\cdots \slashed{q}_{V-1}\Gamma_{n_{\sigma_V}}^{\mu_{\sigma_V}}~^]~\Bigr)^{\alpha_3}_{~~\alpha_4}
\mathcal{I}^{\tiny\mbox{dim-2}}_{\boldsymbol{\sigma}}~.
\end{equation}
In order to transform these fermion chains to the standard forms, we first move all $\slashed{\ell}_i$ and $\slashed{q}_i$ to the right side of $\Gamma$'s. Since the product of gamma matrices in each fermion chain are anti-symmetrized, this operation only produces some extra minus signs,
\begin{eqnarray}
\mathcal{F}_{\tiny\mbox{step-0}}&=&i^{2V-2}\prod_{j=1}^V\frac{-i(-1)^{(j-1)(n_j+n_{\sigma_j})}g_{n_j}}{n_j!}\\
&&~~~~~~~~~~~~~~~~\times\Bigl(~^[~\Gamma_{n_1}^{\mu_1}\cdots \Gamma_{n_V}^{\mu_V}\slashed{\ell}_1\cdots \slashed{\ell}_{V-1}~^]~\Bigr)^{\alpha_1}_{~~\alpha_2}\times \Bigl(~^[~\Gamma_{n_{\sigma_1}}^{\mu_{\sigma_1}}\cdots
\Gamma_{n_{\sigma_V}}^{\mu_{\sigma_V}}
\slashed{q}_1\cdots \slashed{q}_{V-1}~^]~\Bigr)^{\alpha_3}_{~~\alpha_4}\mathcal{I}^{\tiny\mbox{dim-2}}_{\boldsymbol{\sigma}}~.\nonumber
\end{eqnarray}
Following the same operation as in \S\ref{subsec:7loopIntegral}, we set all external momenta to zero, and rewrite the integrand as,
\begin{eqnarray}
\mathcal{F}_{\tiny\mbox{step-1}}&=& \prod_{j=1}^V\frac{-i(-1)^{(j-1)(n_j+n_{\sigma_j})}g_{n_j}}{n_j!}
\mbox{sign}(\boldsymbol{\sigma})\\
&&~~~~~~~~~~~~~~~~\times\Bigl(~^[~\Gamma_{n_1}^{\mu_1}\cdots \Gamma_{n_V}^{\mu_V}\slashed{\ell}_1\cdots \slashed{\ell}_{V-1}~^]~\Bigr)^{\alpha_1}_{~~\alpha_2}\times \Bigl(~^[~\Gamma_{n_{\sigma_1}}^{\mu_{\sigma_1}}\cdots
\Gamma_{n_{\sigma_V}}^{\mu_{\sigma_V}}
\slashed{\ell}_1\cdots \slashed{\ell}_{V-1}~^]~\Bigr)^{\alpha_3}_{~~\alpha_4}\mathcal{I}^{\tiny\mbox{dim-2}}_{\boldsymbol{\sigma}}~.\nonumber
\end{eqnarray}
After PV reduction and dimensional shifting, we further get
\begin{eqnarray}
\mathcal{F}_{\tiny\mbox{step-2,3}}&=& \prod_{j=1}^V\frac{-i(-1)^{(j-1)(n_j+n_{\sigma_j})}g_{n_j}}{n_j!}
\frac{\mbox{sign}(\boldsymbol{\sigma})}{2^{V-1}}\\
&&~~~~~~~~~~~~~~~~~~~~~~~~~~~~~\times\Bigl(~^[~\Gamma_{n_1}^{\mu_1}\cdots \Gamma_{n_V}^{\mu_V}\Gamma_{V-1}^{\nu}~^]~\Bigr)^{\alpha_1}_{~~\alpha_2}\times \Bigl(~^[~\Gamma_{n_{\sigma_1}}^{\mu_{\sigma_1}}\cdots
\Gamma_{n_{\sigma_V}}^{\mu_{\sigma_V}}\Gamma_{V-1}^{\nu}
~^]~\Bigr)^{\alpha_3}_{~~\alpha_4}\mathcal{I}^{\tiny\mbox{dim-4}}_{\boldsymbol{\sigma}}~,\nonumber
\end{eqnarray}
where the two-dimensional tensor integral has been converted to four-dimensional scalar integral, without altering the UV counterterms. Finally, we want to rearrange the gamma matrices in the second fermion chain in the ordering of the first fermion chain. The factor which arises during the rearrangement will be denoted by $\mbox{sign}(\boldsymbol{\sigma}_{\Gamma})$, {\sl i.e.},
\begin{equation}
\Bigl(~^[~\Gamma_{n_{\sigma_1}}^{\mu_{\sigma_1}}\cdots
\Gamma_{n_{\sigma_V}}^{\mu_{\sigma_V}}\Gamma_{V-1}^{\nu}
~^]~\Bigr)^{\alpha_3}_{~~\alpha_4}
=\mbox{sign}(\boldsymbol{\sigma}_{\Gamma})
\Bigl(~^[~\Gamma_{n_1}^{\mu_1}\cdots
\Gamma_{n_V}^{\mu_V}\Gamma_{V-1}^{\nu}~^]~\Bigr)^{\alpha_3}_{~~\alpha_4}~.
\end{equation}
Then we obtain the final integrand in terms of the standard anti-symmetric fermion chains as,
\begin{equation}\label{eqn:fermionChainIntegrandGeneral}
\mathcal{I}^{(V-1)}[n_1,\ldots,n_V]= \prod_{j=1}^V\frac{-i(-1)^{(j-1)(n_j+n_{\sigma_j})}g_{n_j}}{n_j!}
\frac{\mbox{sign}(\boldsymbol{\sigma})\mbox{sign}(\boldsymbol{\sigma}_{\Gamma})}{2^{V-1}}\delta_{i_1}^{i_2}\delta_{i_3}^{i_4}
(\gamma^{\mu_1\cdots \mu_n})^{\alpha_1}_{~~\alpha_2}
 (\gamma_{\mu_1\cdots \mu_n})^{\alpha_3}_{~~\alpha_4}\mathcal{I}^{\tiny\mbox{dim-4}}_{\boldsymbol{\sigma}}~,
\end{equation}
in which $n=V-1+\sum_{i=1}^V n_i$. Note that in order to obtain the complete integrand, we need to sum over all different permutations of $\{n_i\}$ in the first fermion chain, and also sum over all $\boldsymbol{\sigma}\in S_V$ in the second fermion chain. The UV divergence is contained in the integrand $\mathcal{I}^{\tiny\mbox{dim-4}}_{\boldsymbol{\sigma}}$, and again by using the large momentum expansion and the graphical function method with \verb!TwoPoint! command, we can get the results of the integrand $\mathcal{I}^{\tiny\mbox{dim-4}}_{\boldsymbol{\sigma}}$ and the corresponding UV divergence.

\subsection{The beta functions of evanescent operators}
\label{subsec:7loopBeta}

The $Z$-factor of $g_n$ can be evaluated by summing over the contribution of all diagrams discussed in eqn.(\ref{eqn:fermionChainIntegrand}) and (\ref{eqn:fermionChainIntegrandGeneral}). The beta functions can be extracted from the $Z$-factors by the traditional strategy following \cite{Bondi:1988fp,Bondi:1989nq}. However
here we would like to adopt an alternative way to derive the beta functions. We take the convention of $\overline{\mbox{MS}}$ scheme in $(2-2\epsilon)$-dimension, and follow the definition of beta functions as,
\begin{equation}\label{eqn:defBeta}
g_i^{\tiny\mbox{bare}}=Z_i\widetilde{\mu}^{2\epsilon}g_i~~~,~~~\beta(g_i)=2\epsilon g_i+\frac{\partial g_i}{\partial \ln {\mu}}~~,~~i=0,1,2,\ldots
\end{equation}
It is convenient to define the quantity $Y_n=(Z_n-1)g_n$, and by direct evaluation using relation (\ref{eqn:defBeta}) and $\frac{\partial \widetilde{\mu}^{\alpha \epsilon}}{\partial \ln {\mu}}=\alpha \epsilon \widetilde{\mu}^{\alpha \epsilon}$, we get an alternative expression for evaluating beta functions as,
\begin{equation}\label{eqn:defBeta-Y}
\beta(g_n)=-2\epsilon Y_n-\frac{\partial Y_n}{\partial \ln {\mu}}
=-2\epsilon Y_n-\frac{\partial Y_n}{\partial g_0}\frac{\partial g_0}{\partial\ln{\mu}}-\sum_{i=1,2,\ldots}\frac{\partial Y_n}{\partial g_i}\frac{\partial g_i}{\partial\ln{\mu}}~.
\end{equation}
Let us define $g_0=g$, and focus on the renormalization group flow around $\{{\rm{g}}\}/ g_0=0$. The beta function (\ref{eqn:defBeta-Y}) can be explicitly evaluated by
\begin{equation}\label{eqn:defBeta-compute}
\beta(g_n)=-2\epsilon Y_n-\frac{\partial Y_n}{\partial g}\Big(-2\epsilon g+\beta(g)\Big)-\sum_{i=1,2,\ldots}\frac{\partial Y_n}{\partial g_i}\Big(-2\epsilon g_i+\beta(g_i)\Big)~.
\end{equation}
However, in order to obtain the correct beta functions, one cannot trivially set $\{{\rm{g}}\}/g_0=0$ in $Y_n$, because $Y_n$ may contain terms like $g_i f(g,g_1,g_2,\ldots)$, which gives non-zero contribution to the beta function,
\begin{equation}
\left.\frac{\partial }{\partial g_i}\Big(g_i f(g,g_1,g_2,\ldots)\Big)\right|_{\{{\rm{g}}\}/g_0=0}
= f(g,0,\ldots)~.
\end{equation}
Therefore, it is only allowable to set $\{{\rm{g}}\}/g_0=0$ after $\beta(g_n)$ in (\ref{eqn:defBeta-compute}) has been obtained. We can define $\beta_0(g_n):=\beta(g_n)\bigr|_{\{{\rm{g}}\}/g_0=0}$, and compute the beta function around $\{{\rm{g}}\}/g_0=0$ as,
\begin{equation}
\beta_0(g_n)
=\left.\left(-2\epsilon Y_n+2\epsilon g\frac{\partial Y_n}{\partial g}-\sum_{i=1,2,\ldots}\frac{\partial Y_n}{\partial g_i}\beta_0(g_i)\right)\right|_{\{{\rm{g}}\}/g_0=0}~,
\end{equation}
where we have ignored the $\mathcal{O}(g^{n+2})$  order corrections.

We have used \verb!TwoPoint! command in \verb!HyperlogProcedures! package to compute all relevant UV divergences in eqn.(\ref{eqn:fermionChainIntegrand}) and (\ref{eqn:fermionChainIntegrandGeneral}), and evaluated the beta functions by above mentioned formula. Here we present the corresponding results.

The $Y_3, Y_4$ and $Y_7$  are given by,
\begin{eqnarray}
&&Y_3=\frac{3\zeta_3-4}{\epsilon}g^4~~~,~~~Y_4=-\frac{6(3\zeta_3-4)}{\epsilon^2}g^5
+\frac{1}{\epsilon}\left(-15\zeta_5+\frac{\pi^4}{10}+54\zeta_3-\frac{107}{2}\right)g^5
-\frac{8}{\epsilon}gg_3~,\\
&&Y_7=-\frac{140}{\epsilon }g_3^{2}+\frac{140(3\zeta_{3}-4)}{\epsilon }g^{3}g_4+\left(-\frac{490(3\zeta_{3}-4)}{\epsilon^{2}}
-\frac{21(65+\pi^{4}-340\zeta_{3}+250\zeta_{5})}{2\epsilon }\right)g^{4}g_3\nonumber\\
&&~~~~~~~~~~~~~~~~~~~~+\left(-\frac{210(3\zeta_{3}-4)^{2}}{\epsilon ^{3}}+\frac{(3\zeta_{3}-4)(7\pi^{4}+11700\zeta_{3}-4650\zeta_{5}-9145)}{2\epsilon ^{2}}
+\frac{y_{71}}{\epsilon}\right)g^{8}~,
\end{eqnarray}
in which,
\begin{eqnarray*}
y_{71}&=&-70465+129855\zeta_{3}-\frac{161\pi^{4}}{2}
-\frac{365955\zeta_{5}}{4}+\frac{50\pi^{6}}{21}-52020\zeta_{3}^{2}
+\frac{57\pi^{4}\zeta_{3}}{2}-14175\zeta_{7}-\frac{3083\pi^{8}}{140}\\
&&+\frac{717075\zeta_{3}\zeta_{5}}{2}+87480\zeta_{53}-\frac{25\pi^{6}\zeta_{3}}{14}
+199395\zeta_{3}^{3}+15\pi^{4}\zeta_{5}-\frac{1863855\zeta_{9}}{4}
-\frac{641925\zeta_{5}^{2}}{2}+\frac{1533735\zeta_{3}\zeta_{7}}{8}\\
&&+\frac{22761\pi^{6}\zeta_{5}}{28}-\frac{66825\zeta_{3}^{2}\zeta_{5}}{4}
-\frac{204849\pi^{4}\zeta_{7}}{20}-\frac{9218205\pi^{2}\zeta_{9}}{4}+\frac{738993213\zeta_{11}}{32}
-\frac{614547\zeta_{533}}{2}~,
\end{eqnarray*}
where $\zeta_{i_1i_2\cdots i_k}$ is the multivariate zeta function, and for positive integer $i_k$'s it is defined as
\begin{equation*}
\zeta_{i_1i_2\cdots i_k}=\sum_{n_1>n_2>\cdots >n_k>0}\frac{1}{n_1^{i_1}n_2^{i_2}\cdots n_k^{i_k}}~.
\end{equation*}
The corresponding beta functions reads,
\begin{equation}
\beta_0(g_3)=6(3\zeta_3-4) g^4~~~,~~~\beta_0(g_4)=\frac{4}{5}\Big(-150\zeta_5+\pi^4+540\zeta_3-535\Big)g^5~,
\end{equation}
and
\begin{eqnarray}
\beta_0(g_7)&=&\Bigl(-986510+1817970\zeta_{3}-1127\pi^{4}-\frac{2561685\zeta_{5}}{2}
\frac{100\pi^{6}}{3}-728280\zeta_{3}^{2}\nonumber\\
&&~~+399\pi^{4}\zeta_{3}-198450\zeta_{7}-\frac{3083\pi^{8}}{10}+5019525\zeta_{3}\zeta_{5}
+1224720\zeta_{53}-25\pi^{6}\zeta_{3}\nonumber\\
&&~~+2791530\zeta_{3}^{3}+210\pi^{4}\zeta_{5}-\frac{13046985\zeta_{9}}{2}-4493475\zeta_{5}^{2}+\frac{10736145\zeta_{3}\zeta_{7}}{4}
+\frac{22761\pi^{6}\zeta_{5}}{2}\nonumber\\
&&~~~~~~-\frac{467775\zeta_{3}^{2}\zeta_{5}}{2}-\frac{1433943\pi^{4}\zeta_{7}}{10}-\frac{64527435\pi^{2}\zeta_{9}}{2}+\frac{5172952491\zeta_{11}}{16}-4301829\zeta_{533}\Bigr)g^8~.\nonumber
\end{eqnarray}
Our results for $\beta_0(g_3)$ and $\beta_0(g_4)$ are in agreement\footnote{Note that in this paper the definition of $g_n$ is differed by a factor of $n!$. We also absorb the $\frac{1}{4\pi}$ factor into $g_n$.} with \cite{Gracey:2016mio}.

\section{Conclusion}
\label{sec:conclusion}

In this paper, the study of the UV structure in the Gross-Neveu model reveals several critical insights into its renormalizability. By introducing the novel discrete symmetry $\chalf$-parity \eqref{eqn:chalf}, we successfully explain the anomaly arising from the conflict between naive power-counting expectation and the explicit computation about the absence of $O_n$-type UV divergence at $n$-loop. This symmetry, defined through charge conjugation acting on one of the $\bar{\Psi}\Gamma^{(n)}\Psi$ bilinears, imposes a selection rule that suppressing the $\chalf$-odd operators while allowing the $\chalf$-even operators to emerge as counterterms. By the $\chalf$-symmetry (\ref{eqn:statement}), the observations in the literatures, such as the absence of $O_1$ and $O_2$-type divergences at 3 and 4-loops, and the appearance of $O_3, O_4$ in 3 and 4-loops respectively, can be understood. It also predicts the absence of $O_5,O_6$, while the appearance of $O_7,O_8$ at 7 and 8-loops should be confirmed by renormalization computation. This mechanism highlights the importance of discrete symmetries in constraining renormalization group flows, even in the theories with the infinite marginal operators. It is tempting to generalize the $\chalf$-parity analysis to other related models, such as the Gross-Neveu-Yukawa theory or the gauged models, where mixing between the $\chalf$-even and odd operators might occur due to the interactions like Yukawa couplings or gauge fields.

We have successfully performed the 7-loop renormalization computation of the beta function $\beta_0(O_7)$ for the evanescent operator $O_7$. This is the first time the beta function is computed analytically beyond 5-loop in a model with spinning particles. The agreement of the lower-loop results, {\sl e.g.}, $\beta_0(g_3)$ and $\beta_0(g_4)$, with prior work validates the robustness of this approach. The computation of the 7-loop beta function for $O_7$ demonstrates significant technical advancements. The achievement is only possible with the help of the graphical function method together with many traditional techniques. The graphical function is currently the only method to evaluate 7-loop integrals analytically, but by now it was only applied to scalar models. We have successfully extended this method to a model with fermions. With the UV divergence computed by the graphical function method, the anomalous dimensions and the beta functions can be computed to very high loop by combining the OPE and the large momentum expansion \cite{Huang:2024hsn}. The application of these techniques in the renormalization computations not only helps in resolving technical challenges, but also drives the continuous enhancement and sophistication of these theoretical methods themselves. Certainly, these techniques can be applied to even higher loop orders, and we need more investigation towards this.

Although we have calculated the beta function $\beta_0(g_7)$ of the operator $O_7$ up to the 7-loop order, the results of the beta function $\beta_0(g_0)$  are still only up to the 4-loop order. In order to compute the complete 5-loop beta functions, we need to evaluate 5-loop integrals with numerators, in which case the graphical function method no longer works. However, the large momentum expansion is still applicable to reduce the problem to the evaluation of two-point integrals, which by nullifying some external momenta would simplify the integrals. To solve IBP of these integrals, we might need to seek help from many available packages \cite{Smirnov:2008iw,Maierhofer:2017gsa,Liu:2022chg,Wu:2025aeg}. This requires further investigation.

Since there are similarities among the renormalization computations of NATM, GTM and the Gross-Neveu model, one would expect that the graphical function method together with other techniques can also boost the study of NATM and GTM. It is interesting to test the computation efficiency in these theories to high loop order. It is also interesting to consider the $O(N)$ Gross-Neveu model where the fermions are Majorana fermions. This will bring differences to the operations of gamma matrix as well as the Feynman diagrams, but not affect the applications of the large momentum expansion and the graphical function method. We hope the future computations for these theories will help to confirm the capability of the method. Another comment concerns the $O(N)$ quartic scalar theory, since the beta function of Gross-Neveu model has been computed from integrals of the four-dimensional quartic scalar theory in this paper, and also because the two models have the same internal symmetry, one wonders if there is some deep connections between them. This could be surveyed by more computations.


\section*{Acknowledgments}

We would like to thank Oliver Schnetz for stimulating conversations on the graphical function method and the usage of  HyperlogProcedures package. RH was supported by the National Natural Science Foundation of China (NSFC) with Grant No.11805102.

\appendix

\bibliographystyle{JHEP}
\bibliography{Hbib}

\providecommand{\href}[2]{#2}\begingroup\raggedright\begin{thebibliography}{10}

\bibitem{Gross:1974jv}
D.~J. Gross and A.~Neveu, {\it {Dynamical Symmetry Breaking in Asymptotically
  Free Field Theories}},  {\em Phys. Rev. D} {\bf 10} (1974) 3235.

\bibitem{Zamolodchikov:1978xm}
A.~B. Zamolodchikov and A.~B. Zamolodchikov, {\it {Factorized s Matrices in
  Two-Dimensions as the Exact Solutions of Certain Relativistic Quantum Field
  Models}},  {\em Annals Phys.} {\bf 120} (1979) 253--291.

\bibitem{Witten:1978qu}
E.~Witten, {\it {Chiral Symmetry, the 1/n Expansion, and the SU(N) Thirring
  Model}},  {\em Nucl. Phys. B} {\bf 145} (1978) 110--118.

\bibitem{Moshe:2003xn}
M.~Moshe and J.~Zinn-Justin, {\it {Quantum field theory in the large N limit: A
  Review}},  {\em Phys. Rept.} {\bf 385} (2003) 69--228,
  [\href{http://arxiv.org/abs/hep-th/0306133}{{\tt hep-th/0306133}}].

\bibitem{Giombi:2016ejx}
S.~Giombi, {\it {Higher Spin \textemdash{} CFT Duality}},  in {\em {Theoretical
  Advanced Study Institute in Elementary Particle Physics}: {New Frontiers in
  Fields and Strings}}, pp.~137--214, 2017.
\newblock \href{http://arxiv.org/abs/1607.02967}{{\tt arXiv:1607.02967}}.

\bibitem{Shankar:2017zag}
R.~Shankar, {\em {Quantum Field Theory and Condensed Matter}}.
\newblock Cambridge University Press, 8, 2017.

\bibitem{Dashen:1973nhu}
R.~F. Dashen and Y.~Frishman, {\it {Thirring model with u(n) symmetry - scale
  invariant only for fixed values of a coupling constant}},  {\em Phys. Lett.
  B} {\bf 46} (1973) 439--442.

\bibitem{Dashen:1974hp}
R.~F. Dashen and Y.~Frishman, {\it {Four Fermion Interactions and Scale
  Invariance}},  {\em Phys. Rev. D} {\bf 11} (1975) 2781.

\bibitem{Bondi:1988fp}
A.~Bondi, G.~Curci, G.~Paffuti, and P.~Rossi, {\it {Ultraviolet Properties of
  the Generalized Thirring Model With U($N$) Symmetry}},  {\em Phys. Lett. B}
  {\bf 216} (1989) 345--348.

\bibitem{Bondi:1989nq}
A.~Bondi, G.~Curci, G.~Paffuti, and P.~Rossi, {\it {Metric and Central Charge
  in the Perturbative Approach to Two-dimensional Fermionic Models}},  {\em
  Annals Phys.} {\bf 199} (1990) 268.

\bibitem{Wetzel:1984nw}
W.~Wetzel, {\it {Two Loop Beta Function for the {Gross-Neveu} Model}},  {\em
  Phys. Lett. B} {\bf 153} (1985) 297--299.

\bibitem{Destri:1988vb}
C.~Destri, {\it {Two Loop Beta Function for Generalized Nonabelian Thirring
  Models}},  {\em Phys. Lett. B} {\bf 210} (1988) 173. [Erratum: Phys.Lett.B
  213, 565 (1988)].

\bibitem{Gracey:1990sx}
J.~A. Gracey, {\it {Three loop calculations in the O(N) Gross-Neveu model}},
  {\em Nucl. Phys. B} {\bf 341} (1990) 403--418.

\bibitem{Tracas:1989hi}
N.~D. Tracas and N.~D. Vlachos, {\it {Three Loop Calculation of the Beta
  Function for the {Gross-Neveu} Model}},  {\em Phys. Lett. B} {\bf 236} (1990)
  333.

\bibitem{Tracas:1990wc}
N.~D. Tracas and N.~D. Vlachos, {\it {Analytic Results for the Gross-Neveu
  Model Beta Function in Three Loops}},  {\em Phys. Rev. D} {\bf 43} (1991)
  3447--3454.

\bibitem{Gracey:1991vy}
J.~A. Gracey, {\it {Computation of the three loop Beta function of the O(N)
  Gross-Neveu model in minimal subtraction}},  {\em Nucl. Phys. B} {\bf 367}
  (1991) 657--674.

\bibitem{Luperini:1991sv}
C.~Luperini and P.~Rossi, {\it {Three loop Beta function(s) and effective
  potential in the Gross-Neveu model}},  {\em Annals Phys.} {\bf 212} (1991)
  371--401.

\bibitem{Tracas:1990kw}
N.~D. Tracas and N.~D. Vlachos, {\it {Higher order calculations and
  renormalization scheme dependence in the 2-d Gross-Neveu model}},  {\em Phys.
  Lett. B} {\bf 257} (1991) 140--144.

\bibitem{Bennett:1999he}
J.~F. Bennett and J.~A. Gracey, {\it {Three loop renormalization of the
  SU(N(c)) nonAbelian Thirring model}},  {\em Nucl. Phys. B} {\bf 563} (1999)
  390--436, [\href{http://arxiv.org/abs/hep-th/9909046}{{\tt hep-th/9909046}}].

\bibitem{Vasiliev:1995qj}
A.~N. Vasiliev, S.~E. Derkachov, and N.~A. Kivel, {\it {A Technique for
  calculating the gamma matrix structures of the diagrams of a total four
  fermion interaction with infinite number of vertices in d =
  (2+epsilon)-dimensional regularization}},  {\em Theor. Math. Phys.} {\bf 103}
  (1995) 487--495.

\bibitem{Vasiliev:1996rd}
A.~N. Vasiliev, M.~I. Vyazovsky, S.~E. Derkachov, and N.~A. Kivel, {\it {On the
  equivalence of renormalizations in standard and dimensional regularizations
  of 2-D four-fermion interactions}},  {\em Theor. Math. Phys.} {\bf 107}
  (1996) 441--455.

\bibitem{Vasiliev:1996nx}
A.~N. Vasiliev, M.~I. Vyazovsky, S.~E. Derkachov, and N.~A. Kivel, {\it
  {Three-loop calculation of the anomalous field dimension in the full
  four-fermion U(N)-symmetric model}},  {\em Teor. Mat. Fiz.} {\bf 107N3}
  (1996) 359--371.

\bibitem{Vasiliev:1997sk}
A.~N. Vasiliev and M.~I. Vyazovsky, {\it {Proof of the absence of
  multiplicative renormalizability of the Gross-Neveu model in the dimensional
  regularization d = 2+2epsilon}},  {\em Theor. Math. Phys.} {\bf 113} (1997)
  1277--1288.

\bibitem{Ali:2001he}
D.~B. Ali and J.~A. Gracey, {\it {Four loop wave function renormalization in
  the nonAbelian Thirring model}},  {\em Nucl. Phys. B} {\bf 605} (2001)
  337--364, [\href{http://arxiv.org/abs/hep-th/0104207}{{\tt hep-th/0104207}}].

\bibitem{Ludwig:2002fu}
A.~W.~W. Ludwig and K.~J. Wiese, {\it {The Four loop beta function in the 2-D
  nonAbelian Thirring model, and comparison with its conjectured 'exact'
  form}},  {\em Nucl. Phys. B} {\bf 661} (2003) 577--607,
  [\href{http://arxiv.org/abs/cond-mat/0211531}{{\tt cond-mat/0211531}}].

\bibitem{Gracey:2008mf}
J.~A. Gracey, {\it {Four loop MS-bar mass anomalous dimension in the
  Gross-Neveu model}},  {\em Nucl. Phys. B} {\bf 802} (2008) 330--350,
  [\href{http://arxiv.org/abs/0804.1241}{{\tt arXiv:0804.1241}}].

\bibitem{Gracey:2016mio}
J.~A. Gracey, T.~Luthe, and Y.~Schroder, {\it {Four loop renormalization of the
  Gross-Neveu model}},  {\em Phys. Rev. D} {\bf 94} (2016), no.~12 125028,
  [\href{http://arxiv.org/abs/1609.05071}{{\tt arXiv:1609.05071}}].

\bibitem{Dugan:1990df}
M.~J. Dugan and B.~Grinstein, {\it {On the vanishing of evanescent operators}},
   {\em Phys. Lett. B} {\bf 256} (1991) 239--244.

\bibitem{Herrlich:1994kh}
S.~Herrlich and U.~Nierste, {\it {Evanescent operators, scheme dependences and
  double insertions}},  {\em Nucl. Phys. B} {\bf 455} (1995) 39--58,
  [\href{http://arxiv.org/abs/hep-ph/9412375}{{\tt hep-ph/9412375}}].

\bibitem{Jin:2022ivc}
Q.~Jin, K.~Ren, G.~Yang, and R.~Yu, {\it {Gluonic evanescent operators:
  classification and one-loop renormalization}},  {\em JHEP} {\bf 08} (2022)
  141, [\href{http://arxiv.org/abs/2202.08285}{{\tt arXiv:2202.08285}}].

\bibitem{Jin:2022qjc}
Q.~Jin, K.~Ren, G.~Yang, and R.~Yu, {\it {Gluonic evanescent operators:
  two-loop anomalous dimensions}},  {\em JHEP} {\bf 02} (2023) 039,
  [\href{http://arxiv.org/abs/2208.08976}{{\tt arXiv:2208.08976}}].

\bibitem{Schnetz:2013hqa}
O.~Schnetz, {\it {Graphical functions and single-valued multiple
  polylogarithms}},  {\em Commun. Num. Theor. Phys.} {\bf 08} (2014) 589--675,
  [\href{http://arxiv.org/abs/1302.6445}{{\tt arXiv:1302.6445}}].

\bibitem{Golz:2015rea}
M.~Golz, E.~Panzer, and O.~Schnetz, {\it {Graphical functions in parametric
  space}},  {\em Lett. Math. Phys.} {\bf 107} (2017), no.~6 1177--1192,
  [\href{http://arxiv.org/abs/1509.07296}{{\tt arXiv:1509.07296}}].

\bibitem{Borinsky:2021gkd}
M.~Borinsky and O.~Schnetz, {\it {Graphical functions in even dimensions}},
  {\em Commun. Num. Theor. Phys.} {\bf 16} (2022), no.~3 515--614,
  [\href{http://arxiv.org/abs/2105.05015}{{\tt arXiv:2105.05015}}].

\bibitem{Schnetz:2024qqt}
O.~Schnetz and S.~Theil, {\it {Notes on graphical functions with numerator
  structure}},  {\em PoS} {\bf LL2024} (2024) 026,
  [\href{http://arxiv.org/abs/2407.17133}{{\tt arXiv:2407.17133}}].

\bibitem{HP}
O.~Schnetz, {\em {\verb!HyperlogProcedures ver. 0.7!}}
\newblock Maple package available on the homepage of the author,
  https://www.math.fau.de/person/oliver-schnetz.

\bibitem{Schnetz:2022nsc}
O.~Schnetz, {\it {\ensuremath{\phi}4 theory at seven loops}},  {\em Phys. Rev.
  D} {\bf 107} (2023), no.~3 036002,
  [\href{http://arxiv.org/abs/2212.03663}{{\tt arXiv:2212.03663}}].

\bibitem{Huang:2024hsn}
R.~Huang, Q.~Jin, and Y.~Li, {\it {From Operator Product Expansion to Anomalous
  Dimensions}},  \href{http://arxiv.org/abs/2410.03283}{{\tt
  arXiv:2410.03283}}.

\bibitem{Tarasov:1996br}
O.~V. Tarasov, {\it {Connection between Feynman integrals having different
  values of the space-time dimension}},  {\em Phys. Rev. D} {\bf 54} (1996)
  6479--6490, [\href{http://arxiv.org/abs/hep-th/9606018}{{\tt
  hep-th/9606018}}].

\bibitem{Tarasov:1997kx}
O.~V. Tarasov, {\it {Generalized recurrence relations for two loop propagator
  integrals with arbitrary masses}},  {\em Nucl. Phys. B} {\bf 502} (1997)
  455--482, [\href{http://arxiv.org/abs/hep-ph/9703319}{{\tt hep-ph/9703319}}].

\bibitem{Ji:2018yaf}
Y.~Ji and M.~Kelly, {\it {Unitarity violation in noninteger dimensional
  Gross-Neveu-Yukawa model}},  {\em Phys. Rev. D} {\bf 97} (2018), no.~10
  105004, [\href{http://arxiv.org/abs/1802.03222}{{\tt arXiv:1802.03222}}].

\bibitem{Passarino:1978jh}
G.~Passarino and M.~J.~G. Veltman, {\it {One Loop Corrections for e+ e-
  Annihilation Into mu+ mu- in the Weinberg Model}},  {\em Nucl. Phys. B} {\bf
  160} (1979) 151--207.

\bibitem{Smirnov:2008iw}
A.~V. Smirnov, {\it {Algorithm FIRE -- Feynman Integral REduction}},  {\em
  JHEP} {\bf 10} (2008) 107, [\href{http://arxiv.org/abs/0807.3243}{{\tt
  arXiv:0807.3243}}].

\bibitem{Maierhofer:2017gsa}
P.~Maierh\"ofer, J.~Usovitsch, and P.~Uwer, {\it {Kira\textemdash{}A Feynman
  integral reduction program}},  {\em Comput. Phys. Commun.} {\bf 230} (2018)
  99--112, [\href{http://arxiv.org/abs/1705.05610}{{\tt arXiv:1705.05610}}].

\bibitem{Wu:2023upw}
Z.~Wu, J.~Boehm, R.~Ma, H.~Xu, and Y.~Zhang, {\it {NeatIBP 1.0, a package
  generating small-size integration-by-parts relations for Feynman integrals}},
   {\em Comput. Phys. Commun.} {\bf 295} (2024) 108999,
  [\href{http://arxiv.org/abs/2305.08783}{{\tt arXiv:2305.08783}}].

\bibitem{Vladimirov:1979zm}
A.~A. Vladimirov, {\it {Method for Computing Renormalization Group Functions in
  Dimensional Renormalization Scheme}},  {\em Theor. Math. Phys.} {\bf 43}
  (1980) 417.

\bibitem{Chetyrkin:1980pr}
K.~G. Chetyrkin, A.~L. Kataev, and F.~V. Tkachov, {\it {New Approach to
  Evaluation of Multiloop Feynman Integrals: The Gegenbauer Polynomial x Space
  Technique}},  {\em Nucl. Phys. B} {\bf 174} (1980) 345--377.

\bibitem{Caswell:1981ek}
W.~E. Caswell and A.~D. Kennedy, {\it {A Simple Approach to Renormalization
  Theory}},  {\em Phys. Rev. D} {\bf 25} (1982) 392.

\bibitem{Chetyrkin:1982nn}
K.~G. Chetyrkin and F.~V. Tkachov, {\it {Infrared R Operation and Ultraviolet
  Counterterms in the MS Scheme}},  {\em Phys. Lett. B} {\bf 114} (1982)
  340--344.

\bibitem{CHETYRKIN1984419}
K.~Chetyrkin and V.~Smirnov, {\it R*-operation corrected},  {\em Physics
  Letters B} {\bf 144} (1984), no.~5 419--424.

\bibitem{Larin:2002sc}
S.~Larin and P.~van Nieuwenhuizen, {\it {The Infrared R* operation}},
  \href{http://arxiv.org/abs/hep-th/0212315}{{\tt hep-th/0212315}}.

\bibitem{Smirnov:2002pj}
V.~A. Smirnov, {\it {Applied asymptotic expansions in momenta and masses}},
  {\em Springer Tracts Mod. Phys.} {\bf 177} (2002) 1--262.

\bibitem{Liu:2022chg}
X.~Liu and Y.-Q. Ma, {\it {AMFlow: A Mathematica package for Feynman integrals
  computation via auxiliary mass flow}},  {\em Comput. Phys. Commun.} {\bf 283}
  (2023) 108565, [\href{http://arxiv.org/abs/2201.11669}{{\tt
  arXiv:2201.11669}}].

\bibitem{Wu:2025aeg}
Z.~Wu, J.~B\"ohm, R.~Ma, J.~Usovitsch, Y.~Xu, and Y.~Zhang, {\it {Performing
  integration-by-parts reductions using NeatIBP 1.1 + Kira}},
  \href{http://arxiv.org/abs/2502.20778}{{\tt arXiv:2502.20778}}.

\end{thebibliography}\endgroup

\end{document}